\documentclass[reprint,prb,aps,twocolumn,superscriptaddress,10pt]{revtex4-1}
\usepackage{amsmath}
\usepackage[english]{babel}
\usepackage{graphicx}
\usepackage{color}
\usepackage{caption}
\usepackage{ulem}
\usepackage{soul}

\newcommand{\MoSe}{{${\rm MoSe_2}$}}

\newcommand{\moire}{moir\'e}

\newcommand{\Vtg}{$V_{\rm tg}$}
\newcommand{\Vbg}{$V_{\rm bg}$}
\newcommand{\VE}{$V_{\rm E}$}
\newcommand{\Vmu}{$V_{\rm \mu}$}
\newcommand{\DR}{$\Delta R/R_0$}
\addto\captionsenglish{}

\newif\ifdraft
\drafttrue

\def \ETH{Institute for Quantum Electronics, ETH Z\"urich, CH-8093 Z\"urich, Switzerland}

\def \NIMSRCFM{Research Center for Functional Materials,
\\ \normalsize National Institute for Materials Science, Tsukuba, Ibaraki 305-0044, Japan}
\def \NIMSICMN{International Center for Materials Nanoarchitectonics,
\\ \normalsize National Institute for Materials Science, Tsukuba, Ibaraki 305-0044, Japan}
\def \Munich{Department of Physics and Institute for Advanced Study,\\
Technical University of M\"unich, 85748 Garching, Germany}
\def \Munchen{M\"unchen Center for Quantum Science and Technology, Schellingstrasse 4, 80799 M\"unich, Germany}
\def \RIKEN{Center for Emergent Matter Science, RIKEN, 2-1 Hirosawa, Wako, Saitama 351-0198, Japan}
\def \Technion{Physics Department and Solid State Institute, Technion–Israel Institute of Technology, 32000 Haifa, Israel}

\begin{document}
\author{
Ido Schwartz$^{1,2,\dagger}$, Yuya Shimazaki$^{1,3,\dagger}$, Clemens Kuhlenkamp$^{1,4,5}$, Kenji Watanabe$^6$,\\
Takashi Taniguchi$^7$, Martin Kroner$^1$, Ata\c{c} Imamo\u{g}lu$^{1}$\\
~ \\
$^{1}$\ETH\\
$^{2}$\Technion\\
$^{3}$\RIKEN\\
$^{4}$\Munich\\
$^{5}$\Munchen\\
$^{6}$\NIMSRCFM\\
$^{7}$\NIMSICMN \\
~ \\
$^\dagger$These authors contributed equally to this work\\
}

\begin{abstract}
{\bf 
Moire superlattices in twisted transition metal dichalcogenide bilayers have emerged as a rich platform for exploring strong correlations using optical spectroscopy. Despite observation of rich Mott-Wigner physics stemming from an interplay between the periodic potential and Coulomb interactions, the absence of tunnel coupling induced hybridization of electronic states ensured a classical layer degree of freedom in these experiments. Here, we investigate a MoSe$_2$ homobilayer structure where inter-layer coherent tunnelling and layer-selective optical transitions allow for electric field controlled manipulation and measurement of the layer-pseudospin of the ground-state holes.
A striking example of qualitatively new phenomena in this system is our observation of an electrically tunable 2D Feshbach resonance in exciton-hole scattering, which allows us to control the strength of interactions between excitons and holes located in different layers. Our findings enable hitherto  unexplored possibilities for optical investigation of many-body physics, as well as realization of degenerate Bose-Fermi mixtures with tunable interactions, without directly exposing the itinerant fermions to light fields.
}
\end{abstract}

\title{Observation of electrically tunable Feshbach resonances in twisted bilayer semiconductors}
\maketitle


The ability to tune the sign and the strength of interactions between particles is key for realization of quantum simulators.~\cite{Feynman1982} In ultra-cold atom experiments the requisite tunability of interactions is achieved by using Feshbach resonances,~\cite{Feshbach1962,Inouye1998,Chin2010,Bloch2012} which in turn enables many of the impressive developments in the field, ranging from the observation of BEC-to-BCS transition, to the preparation of ultracold molecules.~\cite{Bartenstein2004, Regal2004,Zwierlein2004,Bourdel2004}
Due to their rich internal energy level structure, two cold atoms prepared in a given internal state with total spin $S=S_1$ can be brought into resonance with a bound molecular state of the two atoms with $S \neq S_1$ by applying an external magnetic field. Finite hyperfine coupling allows two atoms in a low energy scattering state (open channel) to hybridize with a bound molecular state (closed channel), resulting in fully tunable scattering phase shifts in the open channel.

Over the last three years, twisted bilayers of two  dimensional (2D) materials have emerged as a new platform for investigating the physics of strongly correlated electrons.~\cite{Cao2018,Sharpe2019,Lu2019,Liu2020,Regan2020,Tang2020,Shimazaki2020,Wang2020} In these synthetic quantum materials partial quenching of kinetic energy in emerging flat bands and strong Coulomb interactions play a central role. The role of optical excitations on the other hand, have been mainly restricted to spectroscopy, revealing features such as incompressibility or emergent charge order of the interacting electron system.~\cite{Regan2020,Tang2020,Shimazaki2020,Shimazaki2021} Enhancing the strength of exciton-electron interactions could drastically enrich the physics of this system by allowing for photo-induced correlated states in degenerate Bose-Fermi mixtures.

Here, we demonstrate an electrically tunable Feshbach resonance in a twisted homobilayer MoSe$_2$ heterostructure and thereby pave the way for achieving tunable interactions between bosonic (exciton or polariton) and fermionic (electron or hole) particles. While the presence of a layer pseudospin degree of freedom allows us to obtain open and closed channels that can be tuned in and out of resonance, coherent hole tunneling between the layers provides the counterpart of hyperfine interaction that leads to hybridization of scattering and bound molecular states.

The interplay between strong Coulomb interactions, the moire superlattice and coherent interlayer hole tunneling is key to understanding the rich physics of the twisted \MoSe /hBN/\MoSe\ heterostructure that we study. To unravel its properties, we use the electric field ($E_z$) and hole density dependence of the exciton-polaron spectrum~\cite{Sidler2016,Efimkin2017}. We find a striking moire site and filling factor dependence of the many-body system: for unity filling of the moire superlattice ($\nu=1$) we observe Mott-like correlated insulator states in the top or bottom layers. More generally for $\nu \le 1$, the hole states show no evidence for layer-hybridization. For $\nu > 1$ on the other hand, the excess holes occupy layer hybridized sites in the moire lattice leading to strong $E_z$ dependence of the associated attractive exciton-polaron resonance energy. Upon further increasing $E_z$, we observe an asymmetric avoided crossing in the optically excited state, demonstrating the hybridisation of the top layer trion state with a continuum of scattering states composed of a top layer exciton and a bottom layer hole.
This observation provides an unequivocal evidence for a 2D Feshbach resonance~\cite{Takemura2014} and the presence of attractive and repulsive polaron branches associated with the emerging inter-layer Feshbach molecule.

\begin{figure*}[h!]
\centering
\includegraphics[width=1\textwidth]{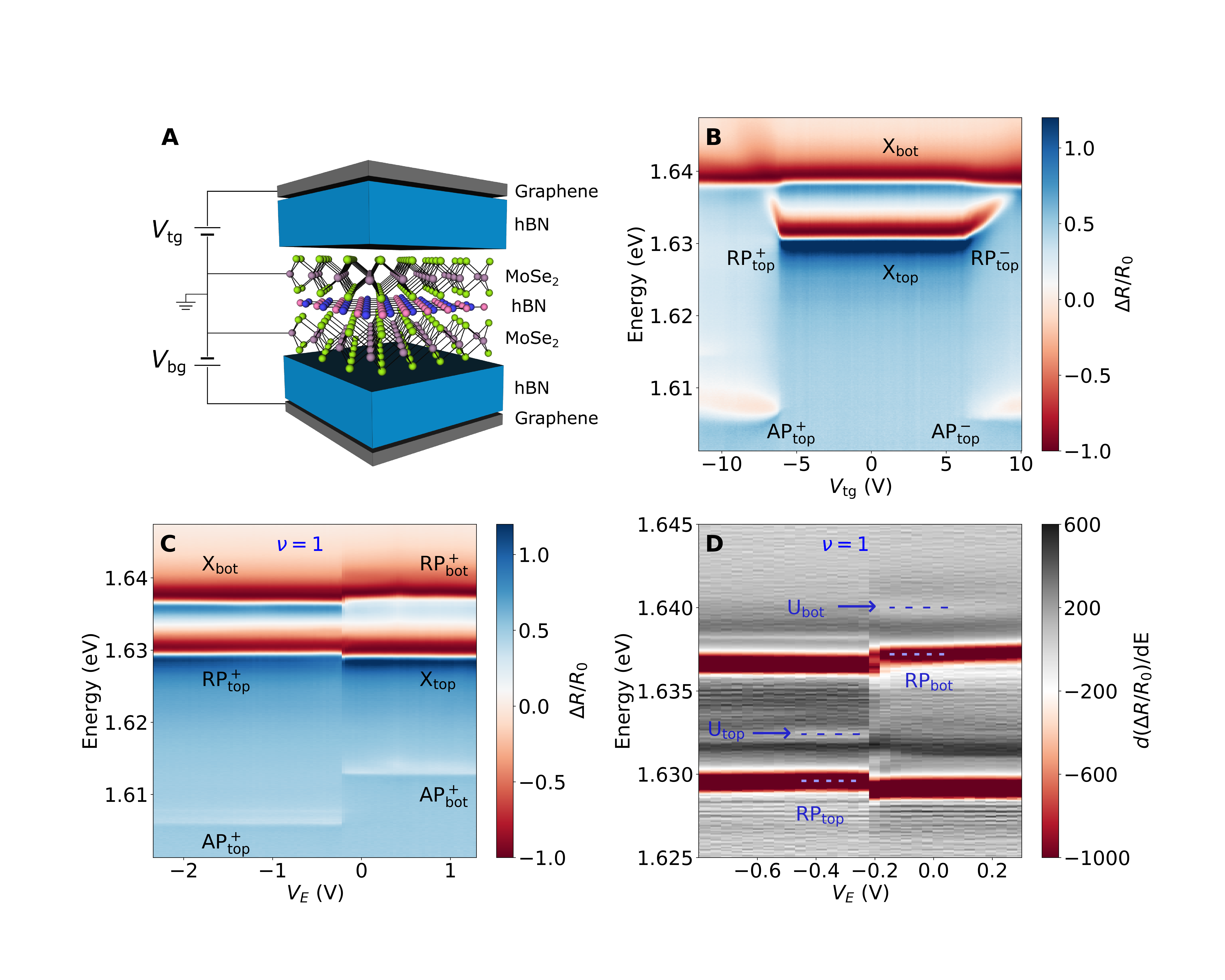}
\caption{{\bf Device structure and basic characteristics.}
        {({\bf A})} Schematic image of the device structure.
        {({\bf B})} Top gate dependent differential reflectance spectra at a fixed back gate voltage, ${\rm V_{BG}=-4}\,V$.
        The top and bottom layer excitons are split due to strain~\cite{Shimazaki2020} and are labeled $\text{X}_{\text{top}}$ and $\text{X}_{\text{bot}}$, respectively.
        Similarly, we use $\text{RP}_{\text{top}}^+$ ($\text{RP}_{\text{top}}^-$) to denote the top layer positive (negative) repulsive polaron, and $\text{AP}_{\text{top}}^+$ ($\text{AP}_{\text{top}}^-$) to denote the top layer attractive polaron.
        {({\bf C})} $V_{\rm E}$ dependent differential reflectance spectra at a fixed chemical potential for unity filling of the moire superlattice $\nu=1$. The bottom layer positive repulsive and attractive polaron resonances are labeled  $\text{RP}_{\text{bot}}^+$ and $\text{AP}_{\text{bot}}^+$, respectively.
        {({\bf D})} $V_{\rm E}$ dependent differential reflectance spectra differentiated with respect to photon energy at a fixed chemical potential for $\nu=1$. The top (bottom) layer umklapp resonance is labeled $\rm U_{top}$ ($\rm U_{bot}$). Blue dashed lines marks the energy of the repulsive polarons and the associated umklapp resonances.
} \label{fig1}
\end{figure*}

Figure~\ref{fig1}A shows a schematic image of the device that we use in our experiment: it consists of two \MoSe\ layers encapsulated between two thick hBN layers and separated by a monolayer hBN tunnel barrier. Top and bottom gate voltages (\Vtg\ and \Vbg , respectively) are applied using transparent thin graphene sheets, while the two TMD layers are grounded.~\cite{Shimazaki2020}
Figure~\ref{fig1}B shows a typical differential reflectance spectra ($\Delta R/R_0 \equiv	(R - R_0)/R_0$) as function of top gate voltage with a fixed back gate voltage of \Vbg$=-4\,\rm V$. Between \Vtg$\approx\pm 6\,\rm V$ the excitation spectrum is dominated by the neutral exciton resonances of the top and bottom layers ($\rm E^0_{X_{top}}=1.632\,eV$ and $\rm E^0_{X_{bot}}=1.64\,eV$, respectively), indicating that both layers are charge-free. By setting $\mathopen| V_{\rm tg}\mathclose| > 6\,\rm V$ the top layer exciton evolves into the repulsive polaron (RP). In addition, an attractive polaron (AP) branch appears at lower energies ($\rm E_{AP_{top}}=1.607\,eV$).~\cite{Sidler2016}

In the following we will use \Vmu\ and \VE\ to denote a linear combination of \Vtg\ and \Vbg\ that results in a minimal change to $E_z$ and the chemical potential $\mu$, respectively (see supplementary material). 
Figure~\ref{fig1}C shows \VE\ dependent \DR\ at $\nu=1$ ($V_\mu=-4.985\,\rm V$). By scanning across $V_E=-0.2\,\rm V$ (corresponding to $E_z = 0$), $\rm AP_{top}$ abruptly disappears in conjunction with the appearance of $\rm AP_{bot}$. 
Similarly to the electron doping regime~\cite{Shimazaki2020}, this abrupt transfer of all holes from one layer to the other indicates a first order phase transition between incompressible Mott-like hole states in the top and bottom layers. This incompressible state results in a periodic potential for excitons and the emergence of an umklapp band.~\cite{Shimazaki2021}
Figure~\ref{fig1}D shows the derivative of \DR\ with respect to energy as a function of \VE\ at $\nu=1$. The umklapp  resonances appear $2.8\,\rm meV$ above the RPs, and are labeled $\rm U_{top}$ and $\rm U_{bot}$.

\begin{figure*}[h!]
\centering
\includegraphics[width=1\textwidth]{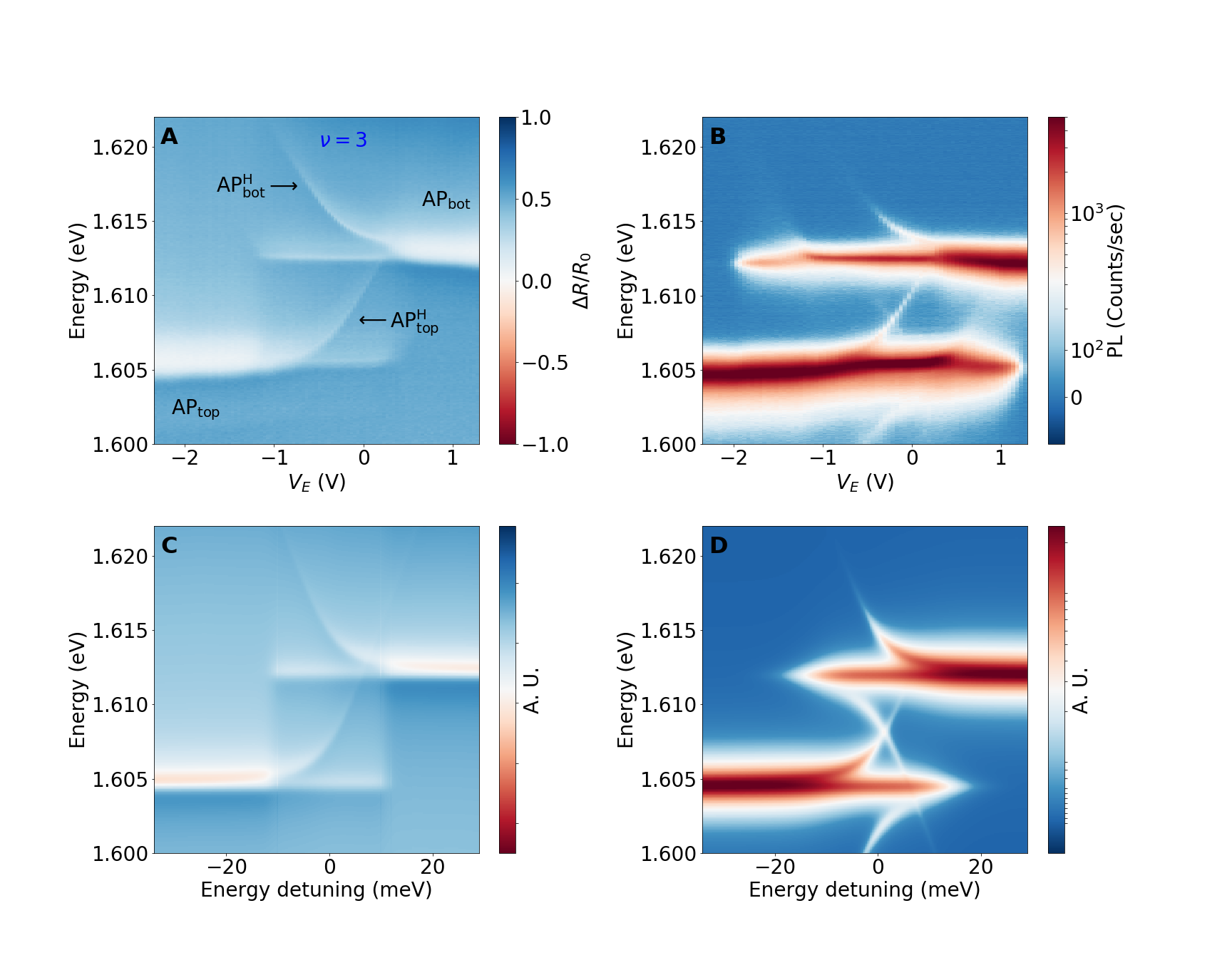}
\caption{{\bf Coherent hole tunneling at $\nu=3$.}
        {({\bf A})} $V_{\rm E}$ dependent differential reflectance spectra at a fixed chemical potential for \moire\ filling factor $\nu=3$ where each lattice site accommodates three holes. Top (bottom) layer AP is labeled $\text{AP}_{\text{top}}$ ($\text{AP}_{\text{bot}}$) and the top (bottom) layer AP originated from hybridized hole state is labeled $\text{AP}^{\rm H}_{\text{top}}$ ($\text{AP}^{\rm H}_{\text{bot}}$)
        {({\bf B})} $V_{\rm E}$ dependent PL spectra at a fixed chemical potential for $\nu=3$.
        {({\bf C,D})} Calculated reflectance spectra (C) and PL (D) at $\nu=3$ as function of energy detuning between top and bottom layers.
}
\label{fig2}
\end{figure*}

Increasing the carrier density to $\nu \ge 1$ reveals a remarkable difference between electron and hole doping: in contrast to changes in electron densities in which a periodic structure of incompressible states appears in the spectra~\cite{Shimazaki2020}, upon increasing the hole density, both $\rm AP_{top}$ and $\rm AP_{bot}$ split and signatures of an avoided crossing appears in one of their branches.
Figure~\ref{fig2}A shows \VE\ dependent \DR\ of AP resonances at $\nu=3$ ($V_\mu=-5.325\,\rm V$). Varying \VE\ from negative to positive, the $\rm AP_{top}$ splits into two branches at $V_E\approx -0.7\, \rm V$, where the energy of one branch remains unchanged and the other branch blue shifts. These observations show that a subset of ground-state holes occupy layer-hybridized states whereas the others are localized within a layer. A complementary picture is observed for the $\rm AP_{bot}$ by scanning \VE\ in the opposite direction. In
Figure~\ref{fig2}B we show the \VE\ dependent photoluminescence (PL) spectra at $\nu=3$, where the avoided crossing structure is fully visible in both $\rm AP_{top}$ and $\rm AP_{bot}$ with a coupling strength of $2t\!\approx\!6\,\rm meV$.
The two radiative decay channels, ensuring the observation of both PL branches in the avoided crossing, provide clear evidence to the existence of hybridized state. The absence of the lower energy branch in \DR\ in turn allows us to conclude that the higher energy AP resonances exhibiting strong \VE\ dependence ($\rm AP_{top}^H$ and $\rm AP_{bot}^H$) originates from the holes occupying the lower energy hybridized state.
Figures~\ref{fig2}C and~\ref{fig2}D shows the calculated spectra as function of energy detuning between the top and bottom layer for reflection and PL, respectively (see supplementary material for the details of the model).

The blue shifting branches of $\rm AP_{top}$ and $\rm AP_{bot}$ have comparable strength for $V_E=-0.2\,\rm V$. Upon increasing (decreasing) $E_z$, we observe that the \DR\ strength of $\rm AP_{top}^H$ ($\rm AP_{bot}^H$) decreases while that of $\rm AP_{bot}^H$ ($\rm AP_{top}^H$) increases. This observation can be understood as arising from the $E_z$ dependence of the  probability amplitude for finding the hybridized hole state in the top or bottom layer: for large and positive (negative) $E_z$, the hole in the hybridized state predominantly resides in the bottom (top) layer. In the language of spins, the Bloch vector of the collective hole layer-pseudospin rotates from the equator to the south (north) pole by increasing (decreasing) $E_z$. Different $\rm AP_{top}$ and $\rm AP_{bot}$ resonance energies on the other hand, allow us to measure the projection of the collective layer-pseudospin of holes occupying the hybridized states.

\begin{figure*}[h!]
\centering
\includegraphics[width=1\textwidth]{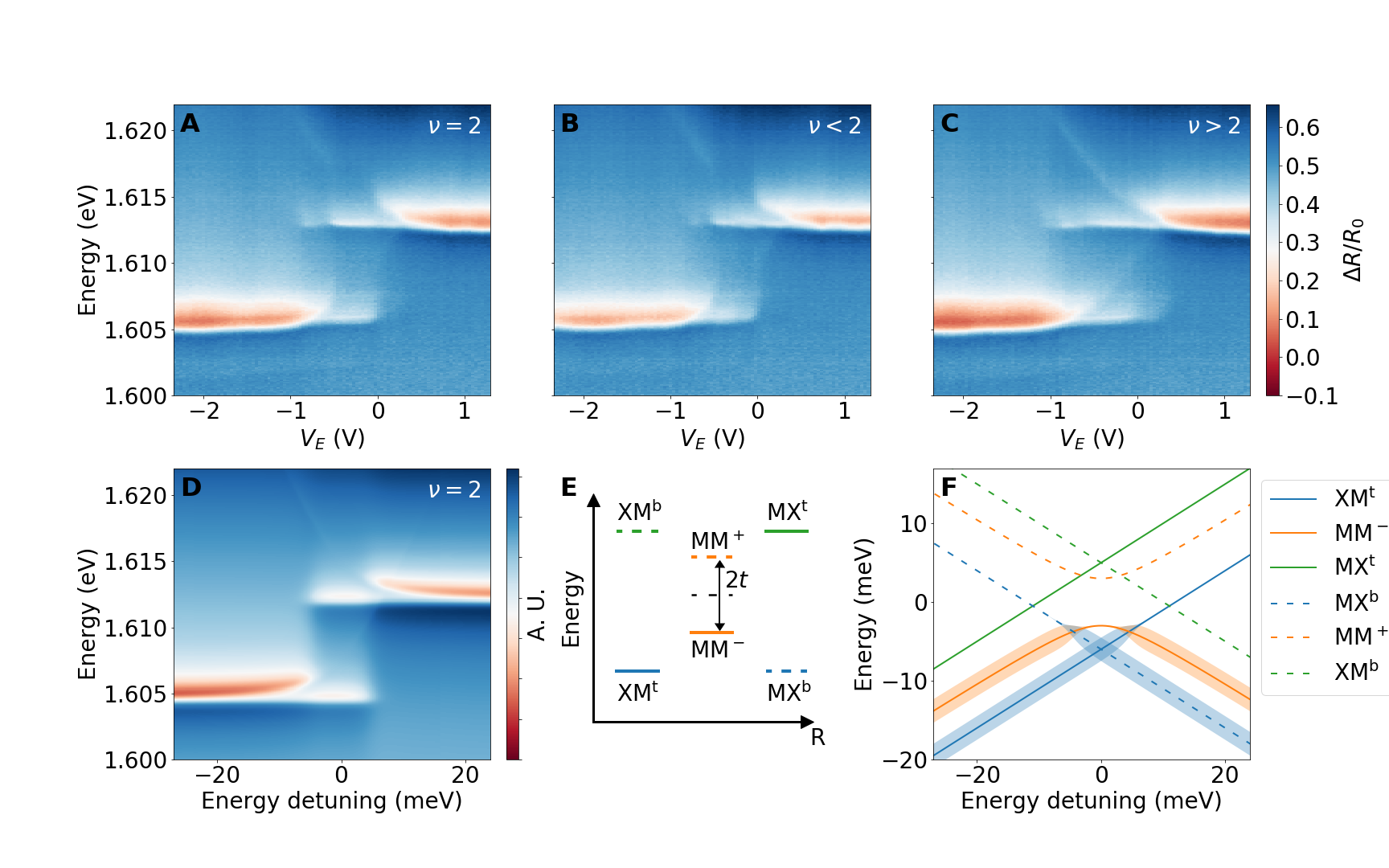}
\caption{{\bf Coherent hole tunneling around filling factor $\nu=2$.}
        {({\bf A-C})} $V_{\rm E}$ dependent differential reflectance spectra at a fixed chemical potential for $\nu=2$ (A), $\nu<2$ (B) and $\nu>2$ (C).
        {({\bf D})} Calculated reflectance spectra at $\nu=2$ as function of energy detuning between top and bottom layers.
        {({\bf E})} Hole energy levels at the three high symmetry points sites $\rm XM$, $\rm MX$ and $\rm MM$ in the \moire\ lattice at zero detuning for both top and bottom layers.
        {({\bf F})} Calculated hole energy levels as function of energy detuning between top and bottom layers. In $\rm XM^t$ ($\rm XM^b$) and $\rm MX^t$ ($\rm MX^b$) the holes occupy the top (bottom) layer, while $\rm MM^-$ and $\rm MM^+$ represent layer hybridized hole states. The width of the semitransparent lines represent hole occupancy at $\nu=2$.
}
\label{fig3}
\end{figure*}

Figure~\ref{fig3}A shows \VE\ dependent \DR\ at $\nu=2$ ($V_\mu=-5.212\,\rm V$). For small detuning between the single-particle states of the two layers ($\rm {-0.5\,V<}V{\rm_ E<0\,V}$), $\rm AP_{top}^H$ and $\rm AP_{bot}^H$ are not visible, indicating that hybridized hole states are not occupied. This gap in the AP spectra is prominent for $1<\nu\le 2$ (Fig.~\ref{fig3}B).
In contrast, $\rm AP_{top}^H$ and $\rm AP_{bot}^H$ resonances can be observed $\forall$\VE\ for $\nu>2$ ($V_\mu=-5.250\,\rm V$), as can be seen in Fig.~\ref{fig3}C.
The emergence of such a gap in the optical excitation spectrum provides a clear evidence that the holes occupy different sites in the \moire\ lattice at large ($V{\rm_ E > 0\,V}$ or $\rm {-0.5\,V>}V{\rm_ E}$) and small ($\rm {-0.5\,V<}V{\rm_ E<0\,V}$) detunings for fixed $\nu$.

The observations reported in Fig.~\ref{fig2} and Fig.~\ref{fig3} can be captured by a simple single-particle model that assumes that holes are subject to a superlattice potential with three minima, strong on-site repulsion and no hopping between the different sites within a layer.
The three minima are attributed to different sites in the \moire\ superlattice for which the local stacking of the metal (M) and chalcogen (X) atoms of the two layers are denoted by $\rm MM$, $\rm MX$ and $\rm XM$.
We conclude from the lack of avoided crossing signatures below $\nu\le 1$ that the lowest single particle \moire\ subbands are located at $\rm XM^t$ ($\rm MX^b$) for $V_E<-0.2\,\rm V$ ($V_E>-0.2\,\rm V$),~\cite{Yu2019} with t{\textbackslash}b indicating top{\textbackslash}bottom layer hole.
When $1<\nu\le 2$ the second \moire\ subband starts to fill: for large energy detuning between the layers this subband is located at the hybridized $\rm MM$ site and for small energy detuning the two lowest energy states are $\rm XM^t$ and $\rm MX^b$.
For $\nu>2$, the third \moire\ subband starts to fill up, leading to the closing of the gap in the avoided crossing in \DR\ and indicating that the $\rm MM$ sites are occupied for all energy detunings between the layers.
Figure~\ref{fig3}E is a schematic illustration of the model showing the energy of the hole states at the three high symmetry sites of the two layers for $E_z = 0$.
Figure~\ref{fig3}F shows the calculated hole energy levels using the simplified model (see supplementary materials).

\begin{figure*}[h!]
\centering
\includegraphics[width=1\textwidth]{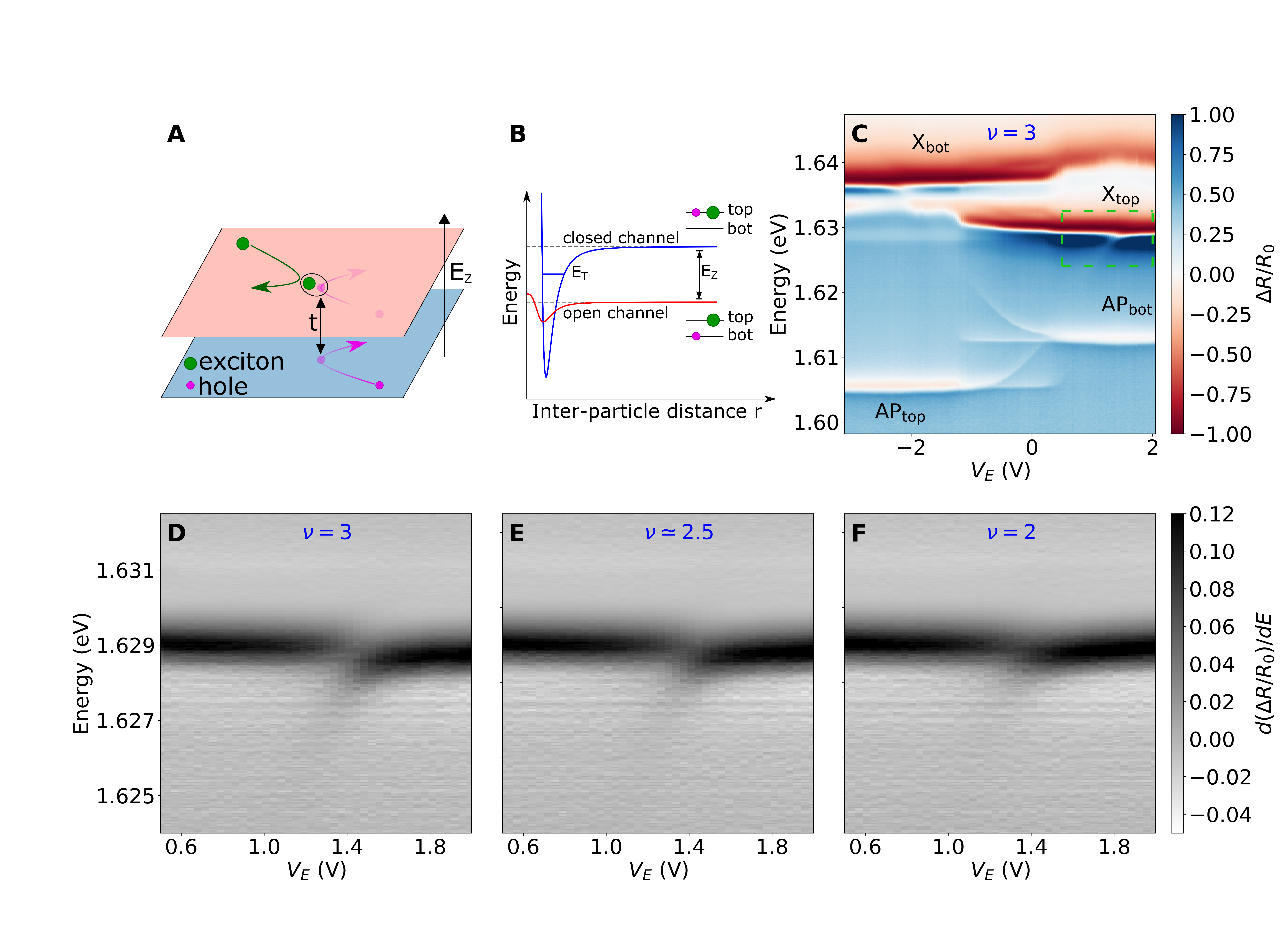}
\caption{{\bf Feshbach resonance.}
        {({\bf A})} Illustration of resonantly enhanced scattering of an exciton and a hole in a bilayer TMD. The Feshbach resonance is achieved by tuning $\rm E_z$ and setting the electrostatic potential difference between the layers to match the intra-layer exciton-hole bound state (trion) energy $\rm E_T$.
        {({\bf B})} Schematic depiction of the potential energy of an exciton and a hole in the open channel (red) and the closed channel (blue).
        {({\bf C})} $V_{\rm E}$ dependent differential reflectance spectra at a fixed chemical potential for $\nu=3$. 
        {({\bf D-F)}} $V_{\rm E}$ dependent differential reflectance spectrum differentiated with respect to energy E, in the area marked with dashed green box in (C) for $\nu=3$, $\nu\simeq 2.5$ and $\nu=2$, respectively. The spectrum around the $\rm X_{top}$ resonance is visibly asymmetric and depends on the density of holes in the opposite (bottom) layer.
} \label{fig4}
\end{figure*}

Having identified the ground-state properties arising from an interplay between Coulomb interactions and inter-layer coherent hole tunneling using optical excitations as a spectroscopic tool, we address the Feshbach physics that emerges in the excited state. 
The relevant scattering process is illustrated in Fig.~\ref{fig4}A. By changing the electric field, the potential energy of holes can be adjusted depending on their layer-configuration. This in turn allows us to adjust the inter and intra-layer exciton-hole potentials with respect to one another, as depicted schematically in Fig.~\ref{fig4}B.
We show a wide range of the \VE\ dependent \DR\ spectra at $\nu=3$, covering both AP and RP resonances in Figure~\ref{fig4}C. 
For concreteness we focus on the $\rm AP_{top}^H$ branch that is associated with holes originally at the $\rm MM$ sites of the moire lattice: as $V_E$ is increased, this branch continues to blue shift while losing oscillator strength. The latter feature is a consequence of the fact that as $V_E$ is increased, the probability of finding a lowest energy $\rm MM$ site hole in the top layer is decreased: since the hole in the initial state of the optical transition is predominantly in the bottom layer, the photon energy required to create a top layer AP increases linearly with $V_E$. Naturally, for large $V_E$ the probability to find the hole in the top layer MM site is small, evidenced by the very faint $\rm AP_{top}$ signal.

Remarkably, for $V_E \ge 1.3\,\rm V$, the oscillator strength of the $V_E$-dependent top layer AP transition starts to increase and exhibits a highly asymmetric anti-crossing with the top-layer exciton transition at $V_E = 1.4\,\rm V$. The corresponding avoided crossing between the bottom layer AP and the bottom layer exciton transitions is observed at $V_E = -2.1\,\rm V$. If this had been a simple zero-dimensional system, such as a quantum-dot molecule, we could have explained the avoided crossing as stemming from hybridization of a top layer trion with a state consisting of a top-layer exciton and a bottom-layer hole.~\cite{Stinaff2006,Krenner2006}
The experiments depicted in Fig.~\ref{fig4}C on the other hand, are carried out in a 2D system with a layer pseudo-spin degree of freedom~\cite{Fertig1989,MacDonald1990}: the resonance responsible for the avoided crossing is between a bound molecular state (trion) and a continuum of scattering states consisting of 2D excitons and holes on $\rm MM$ sites of the moire lattice. This is in stark contrast to the ground-state anti-crossing stemming from inter-layer hybridization of two hole states with the same in-plane momentum. Two important distinctions need to be made: first, due to the fact that we are dealing with particles in 2D, the coupled system supports a bound inter-layer molecular state for all electric fields $V_E$; this molecule originates from intra-layer trions resonantly coupled to a continuum of inter-layer exciton-hole states and is referred to as a Feshbach molecule~\cite{Chin2010}. Second, the observed dependence on hole density and the asymmetry of the avoided crossing can be explained in terms of RP and AP formation associated with this inter-layer molecular state~\cite{Kohstall2012,Schmidt2012}.

Figure~\ref{fig4}D shows the derivative of \DR\ with respect to energy in the spectral region highlighted with the green dashed box in Fig.~\ref{fig4}C: we observe that the splitting is much smaller than the coherent hole tunnel rate. Moreover, the nature of the anti-crossing is strikingly different from that of a simple avoided crossing. The reduction of the splitting is a consequence of the reduced wavefunction overlap between the scattering and molecular states.
Figures~\ref{fig4}E and~\ref{fig4}F are derivatives of \DR\ with respect to energy at $\nu \simeq 2.5$ and $\nu=2$ which show that at lower hole densities the splitting is further reduced: this reduction is consistent with the fact that the oscillator strength of the Feshbach AP resonance as well as the splitting between the AP and RP resonances are many-body effects which scale with the density of holes.
In addition, unlike the ground state, the scattering state has a short radiative lifetime which further suppresses the observed splitting. Last but not least, only holes with a finite probability amplitude to occupy $\rm MM$ sites contribute to the anticrossing and to the strength of the AP resonance.
Our observation of asymmetric reflection spectra is typical for a bound state coupled to a continuum of scattering states~\cite{Chin2010}. These features are in stark contrast to experiments on polariton Feshbach resonances, where signatures of the characteristic scattering physics could not be conclusively demonstrated~\cite{Takemura2014}; there, the cavity-like dispersion of the polaritons leads to a vanishingly small density of states which prohibits an efficient coupling to the continuum.

Our experiments establish that bilayer TMD systems with coherent inter-layer hole tunneling exhibit truly 2D Feshbach resonances that induce inter-layer Feshbach molecules even when the binding energy of the intra-layer trion is much larger than the energy of the scattering state. We use such a resonance to enhance the interactions between excitons and holes residing in different layers by electrically tuning the binding energy of the Feshbach molecule. The interaction strength is valley selective and may be used to induce exciton-mediated ferromagnetic interactions between holes in neighboring $\rm MM$ sites. Consequently, we expect our findings to play a role in exploring quantum magnetism in moire lattices.~\cite{Burch2018} Even though our experiments explore enhancement of exciton-electron scattering, the underlying concept is more general; a remarkable extension of the concepts and techniques we introduce would be provided by an electric field controlled Feshbach resonance between a scattering state of two electrons in a single layer and the bound inter-layer trimer state.~\cite{Slagle2020,Zhang2021}
Furthermore, the experiments that we have reported here establish that combination of applied electric field, inter-layer hole tunneling and layer-selective optical excitations can in principle allow for arbitrary rotation of the layer-pseudo-spin on the Bloch sphere together with projective measurements in the up-down basis. The degree of achievable control could enable a new set of quantum optics experiments in 2D materials, including optical pumping of valley and layer pseudospin and electromagnetically induced transparency exploiting layer coherence.

\vspace{1 cm}

\section*{Acknowledgments}
The authors thank Michael Knap, Richard Schmidt and Puneet A. Murthy for fruitful discussions.
{\bf Funding:} This work was supported by the Swiss National Science Foundation (SNSF) under Grant No. 200021-178909/1. Y.S. acknowledges support from the Japan Society for the Promotion of Science (JSPS) overseas research fellowships. K.W. and T.T. acknowledge support from the Elemental Strategy Initiative conducted by MEXT, Japan, A3 Foresight by JSPS and CREST (grant no. JPMJCR15F3) and JST.

\bibliography{References.bib}

\begin{thebibliography}{31}%
\makeatletter
\providecommand \@ifxundefined [1]{%
 \@ifx{#1\undefined}
}%
\providecommand \@ifnum [1]{%
 \ifnum #1\expandafter \@firstoftwo
 \else \expandafter \@secondoftwo
 \fi
}%
\providecommand \@ifx [1]{%
 \ifx #1\expandafter \@firstoftwo
 \else \expandafter \@secondoftwo
 \fi
}%
\providecommand \natexlab [1]{#1}%
\providecommand \enquote  [1]{``#1''}%
\providecommand \bibnamefont  [1]{#1}%
\providecommand \bibfnamefont [1]{#1}%
\providecommand \citenamefont [1]{#1}%
\providecommand \href@noop [0]{\@secondoftwo}%
\providecommand \href [0]{\begingroup \@sanitize@url \@href}%
\providecommand \@href[1]{\@@startlink{#1}\@@href}%
\providecommand \@@href[1]{\endgroup#1\@@endlink}%
\providecommand \@sanitize@url [0]{\catcode `\\12\catcode `\$12\catcode
  `\&12\catcode `\#12\catcode `\^12\catcode `\_12\catcode `\%12\relax}%
\providecommand \@@startlink[1]{}%
\providecommand \@@endlink[0]{}%
\providecommand \url  [0]{\begingroup\@sanitize@url \@url }%
\providecommand \@url [1]{\endgroup\@href {#1}{\urlprefix }}%
\providecommand \urlprefix  [0]{URL }%
\providecommand \Eprint [0]{\href }%
\providecommand \doibase [0]{http://dx.doi.org/}%
\providecommand \selectlanguage [0]{\@gobble}%
\providecommand \bibinfo  [0]{\@secondoftwo}%
\providecommand \bibfield  [0]{\@secondoftwo}%
\providecommand \translation [1]{[#1]}%
\providecommand \BibitemOpen [0]{}%
\providecommand \bibitemStop [0]{}%
\providecommand \bibitemNoStop [0]{.\EOS\space}%
\providecommand \EOS [0]{\spacefactor3000\relax}%
\providecommand \BibitemShut  [1]{\csname bibitem#1\endcsname}%
\let\auto@bib@innerbib\@empty
\bibitem [{\citenamefont {Feynman}(1982)}]{Feynman1982}%
  \BibitemOpen
  \bibfield  {author} {\bibinfo {author} {\bibfnamefont {R.~P.}\ \bibnamefont
  {Feynman}},\ }\href {\doibase 10.1007/bf02650179} {\bibfield  {journal}
  {\bibinfo  {journal} {International Journal of Theoretical Physics}\ }\textbf
  {\bibinfo {volume} {21}},\ \bibinfo {pages} {467} (\bibinfo {year}
  {1982})}\BibitemShut {NoStop}%
\bibitem [{\citenamefont {Feshbach}(1962)}]{Feshbach1962}%
  \BibitemOpen
  \bibfield  {author} {\bibinfo {author} {\bibfnamefont {H.}~\bibnamefont
  {Feshbach}},\ }\href {\doibase 10.1016/0003-4916(62)90221-x} {\bibfield
  {journal} {\bibinfo  {journal} {Annals of Physics}\ }\textbf {\bibinfo
  {volume} {19}},\ \bibinfo {pages} {287} (\bibinfo {year} {1962})}\BibitemShut
  {NoStop}%
\bibitem [{\citenamefont {Inouye}\ \emph {et~al.}(1998)\citenamefont {Inouye},
  \citenamefont {Andrews}, \citenamefont {Stenger}, \citenamefont {Miesner},
  \citenamefont {Stamper-Kurn},\ and\ \citenamefont {Ketterle}}]{Inouye1998}%
  \BibitemOpen
  \bibfield  {author} {\bibinfo {author} {\bibfnamefont {S.}~\bibnamefont
  {Inouye}}, \bibinfo {author} {\bibfnamefont {M.~R.}\ \bibnamefont {Andrews}},
  \bibinfo {author} {\bibfnamefont {J.}~\bibnamefont {Stenger}}, \bibinfo
  {author} {\bibfnamefont {H.-J.}\ \bibnamefont {Miesner}}, \bibinfo {author}
  {\bibfnamefont {D.~M.}\ \bibnamefont {Stamper-Kurn}}, \ and\ \bibinfo
  {author} {\bibfnamefont {W.}~\bibnamefont {Ketterle}},\ }\href {\doibase
  10.1038/32354} {\bibfield  {journal} {\bibinfo  {journal} {Nature}\ }\textbf
  {\bibinfo {volume} {392}},\ \bibinfo {pages} {151} (\bibinfo {year}
  {1998})}\BibitemShut {NoStop}%
\bibitem [{\citenamefont {Chin}\ \emph {et~al.}(2010)\citenamefont {Chin},
  \citenamefont {Grimm}, \citenamefont {Julienne},\ and\ \citenamefont
  {Tiesinga}}]{Chin2010}%
  \BibitemOpen
  \bibfield  {author} {\bibinfo {author} {\bibfnamefont {C.}~\bibnamefont
  {Chin}}, \bibinfo {author} {\bibfnamefont {R.}~\bibnamefont {Grimm}},
  \bibinfo {author} {\bibfnamefont {P.}~\bibnamefont {Julienne}}, \ and\
  \bibinfo {author} {\bibfnamefont {E.}~\bibnamefont {Tiesinga}},\ }\href
  {\doibase 10.1103/revmodphys.82.1225} {\bibfield  {journal} {\bibinfo
  {journal} {Reviews of Modern Physics}\ }\textbf {\bibinfo {volume} {82}},\
  \bibinfo {pages} {1225} (\bibinfo {year} {2010})}\BibitemShut {NoStop}%
\bibitem [{\citenamefont {Bloch}\ \emph {et~al.}(2012)\citenamefont {Bloch},
  \citenamefont {Dalibard},\ and\ \citenamefont
  {Nascimb{\`{e}}ne}}]{Bloch2012}%
  \BibitemOpen
  \bibfield  {author} {\bibinfo {author} {\bibfnamefont {I.}~\bibnamefont
  {Bloch}}, \bibinfo {author} {\bibfnamefont {J.}~\bibnamefont {Dalibard}}, \
  and\ \bibinfo {author} {\bibfnamefont {S.}~\bibnamefont {Nascimb{\`{e}}ne}},\
  }\href {\doibase 10.1038/nphys2259} {\bibfield  {journal} {\bibinfo
  {journal} {Nature Physics}\ }\textbf {\bibinfo {volume} {8}},\ \bibinfo
  {pages} {267} (\bibinfo {year} {2012})}\BibitemShut {NoStop}%
\bibitem [{\citenamefont {Bartenstein}\ \emph {et~al.}(2004)\citenamefont
  {Bartenstein}, \citenamefont {Altmeyer}, \citenamefont {Riedl}, \citenamefont
  {Jochim}, \citenamefont {Chin}, \citenamefont {Denschlag},\ and\
  \citenamefont {Grimm}}]{Bartenstein2004}%
  \BibitemOpen
  \bibfield  {author} {\bibinfo {author} {\bibfnamefont {M.}~\bibnamefont
  {Bartenstein}}, \bibinfo {author} {\bibfnamefont {A.}~\bibnamefont
  {Altmeyer}}, \bibinfo {author} {\bibfnamefont {S.}~\bibnamefont {Riedl}},
  \bibinfo {author} {\bibfnamefont {S.}~\bibnamefont {Jochim}}, \bibinfo
  {author} {\bibfnamefont {C.}~\bibnamefont {Chin}}, \bibinfo {author}
  {\bibfnamefont {J.~H.}\ \bibnamefont {Denschlag}}, \ and\ \bibinfo {author}
  {\bibfnamefont {R.}~\bibnamefont {Grimm}},\ }\href {\doibase
  10.1103/physrevlett.92.120401} {\bibfield  {journal} {\bibinfo  {journal}
  {Physical Review Letters}\ }\textbf {\bibinfo {volume} {92}},\ \bibinfo
  {pages} {120401} (\bibinfo {year} {2004})}\BibitemShut {NoStop}%
\bibitem [{\citenamefont {Regal}\ \emph {et~al.}(2004)\citenamefont {Regal},
  \citenamefont {Greiner},\ and\ \citenamefont {Jin}}]{Regal2004}%
  \BibitemOpen
  \bibfield  {author} {\bibinfo {author} {\bibfnamefont {C.~A.}\ \bibnamefont
  {Regal}}, \bibinfo {author} {\bibfnamefont {M.}~\bibnamefont {Greiner}}, \
  and\ \bibinfo {author} {\bibfnamefont {D.~S.}\ \bibnamefont {Jin}},\ }\href
  {\doibase 10.1103/physrevlett.92.040403} {\bibfield  {journal} {\bibinfo
  {journal} {Physical Review Letters}\ }\textbf {\bibinfo {volume} {92}},\
  \bibinfo {pages} {040403} (\bibinfo {year} {2004})}\BibitemShut {NoStop}%
\bibitem [{\citenamefont {Zwierlein}\ \emph {et~al.}(2004)\citenamefont
  {Zwierlein}, \citenamefont {Stan}, \citenamefont {Schunck}, \citenamefont
  {Raupach}, \citenamefont {Kerman},\ and\ \citenamefont
  {Ketterle}}]{Zwierlein2004}%
  \BibitemOpen
  \bibfield  {author} {\bibinfo {author} {\bibfnamefont {M.~W.}\ \bibnamefont
  {Zwierlein}}, \bibinfo {author} {\bibfnamefont {C.~A.}\ \bibnamefont {Stan}},
  \bibinfo {author} {\bibfnamefont {C.~H.}\ \bibnamefont {Schunck}}, \bibinfo
  {author} {\bibfnamefont {S.~M.~F.}\ \bibnamefont {Raupach}}, \bibinfo
  {author} {\bibfnamefont {A.~J.}\ \bibnamefont {Kerman}}, \ and\ \bibinfo
  {author} {\bibfnamefont {W.}~\bibnamefont {Ketterle}},\ }\href {\doibase
  10.1103/physrevlett.92.120403} {\bibfield  {journal} {\bibinfo  {journal}
  {Physical Review Letters}\ }\textbf {\bibinfo {volume} {92}},\ \bibinfo
  {pages} {120403} (\bibinfo {year} {2004})}\BibitemShut {NoStop}%
\bibitem [{\citenamefont {Bourdel}\ \emph {et~al.}(2004)\citenamefont
  {Bourdel}, \citenamefont {Khaykovich}, \citenamefont {Cubizolles},
  \citenamefont {Zhang}, \citenamefont {Chevy}, \citenamefont {Teichmann},
  \citenamefont {Tarruell}, \citenamefont {Kokkelmans},\ and\ \citenamefont
  {Salomon}}]{Bourdel2004}%
  \BibitemOpen
  \bibfield  {author} {\bibinfo {author} {\bibfnamefont {T.}~\bibnamefont
  {Bourdel}}, \bibinfo {author} {\bibfnamefont {L.}~\bibnamefont {Khaykovich}},
  \bibinfo {author} {\bibfnamefont {J.}~\bibnamefont {Cubizolles}}, \bibinfo
  {author} {\bibfnamefont {J.}~\bibnamefont {Zhang}}, \bibinfo {author}
  {\bibfnamefont {F.}~\bibnamefont {Chevy}}, \bibinfo {author} {\bibfnamefont
  {M.}~\bibnamefont {Teichmann}}, \bibinfo {author} {\bibfnamefont
  {L.}~\bibnamefont {Tarruell}}, \bibinfo {author} {\bibfnamefont {S.~J. J.
  M.~F.}\ \bibnamefont {Kokkelmans}}, \ and\ \bibinfo {author} {\bibfnamefont
  {C.}~\bibnamefont {Salomon}},\ }\href {\doibase
  10.1103/physrevlett.93.050401} {\bibfield  {journal} {\bibinfo  {journal}
  {Physical Review Letters}\ }\textbf {\bibinfo {volume} {93}},\ \bibinfo
  {pages} {050401} (\bibinfo {year} {2004})}\BibitemShut {NoStop}%
\bibitem [{\citenamefont {Cao}\ \emph {et~al.}(2018)\citenamefont {Cao},
  \citenamefont {Fatemi}, \citenamefont {Demir}, \citenamefont {Fang},
  \citenamefont {Tomarken}, \citenamefont {Luo}, \citenamefont
  {Sanchez-Yamagishi}, \citenamefont {Watanabe}, \citenamefont {Taniguchi},
  \citenamefont {Kaxiras}, \citenamefont {Ashoori},\ and\ \citenamefont
  {Jarillo-Herrero}}]{Cao2018}%
  \BibitemOpen
  \bibfield  {author} {\bibinfo {author} {\bibfnamefont {Y.}~\bibnamefont
  {Cao}}, \bibinfo {author} {\bibfnamefont {V.}~\bibnamefont {Fatemi}},
  \bibinfo {author} {\bibfnamefont {A.}~\bibnamefont {Demir}}, \bibinfo
  {author} {\bibfnamefont {S.}~\bibnamefont {Fang}}, \bibinfo {author}
  {\bibfnamefont {S.~L.}\ \bibnamefont {Tomarken}}, \bibinfo {author}
  {\bibfnamefont {J.~Y.}\ \bibnamefont {Luo}}, \bibinfo {author} {\bibfnamefont
  {J.~D.}\ \bibnamefont {Sanchez-Yamagishi}}, \bibinfo {author} {\bibfnamefont
  {K.}~\bibnamefont {Watanabe}}, \bibinfo {author} {\bibfnamefont
  {T.}~\bibnamefont {Taniguchi}}, \bibinfo {author} {\bibfnamefont
  {E.}~\bibnamefont {Kaxiras}}, \bibinfo {author} {\bibfnamefont {R.~C.}\
  \bibnamefont {Ashoori}}, \ and\ \bibinfo {author} {\bibfnamefont
  {P.}~\bibnamefont {Jarillo-Herrero}},\ }\href {\doibase 10.1038/nature26154}
  {\bibfield  {journal} {\bibinfo  {journal} {Nature}\ }\textbf {\bibinfo
  {volume} {556}},\ \bibinfo {pages} {80} (\bibinfo {year} {2018})}\BibitemShut
  {NoStop}%
\bibitem [{\citenamefont {Sharpe}\ \emph {et~al.}(2019)\citenamefont {Sharpe},
  \citenamefont {Fox}, \citenamefont {Barnard}, \citenamefont {Finney},
  \citenamefont {Watanabe}, \citenamefont {Taniguchi}, \citenamefont
  {Kastner},\ and\ \citenamefont {Goldhaber-Gordon}}]{Sharpe2019}%
  \BibitemOpen
  \bibfield  {author} {\bibinfo {author} {\bibfnamefont {A.~L.}\ \bibnamefont
  {Sharpe}}, \bibinfo {author} {\bibfnamefont {E.~J.}\ \bibnamefont {Fox}},
  \bibinfo {author} {\bibfnamefont {A.~W.}\ \bibnamefont {Barnard}}, \bibinfo
  {author} {\bibfnamefont {J.}~\bibnamefont {Finney}}, \bibinfo {author}
  {\bibfnamefont {K.}~\bibnamefont {Watanabe}}, \bibinfo {author}
  {\bibfnamefont {T.}~\bibnamefont {Taniguchi}}, \bibinfo {author}
  {\bibfnamefont {M.~A.}\ \bibnamefont {Kastner}}, \ and\ \bibinfo {author}
  {\bibfnamefont {D.}~\bibnamefont {Goldhaber-Gordon}},\ }\href {\doibase
  10.1126/science.aaw3780} {\bibfield  {journal} {\bibinfo  {journal}
  {Science}\ }\textbf {\bibinfo {volume} {365}},\ \bibinfo {pages} {605}
  (\bibinfo {year} {2019})}\BibitemShut {NoStop}%
\bibitem [{\citenamefont {Lu}\ \emph {et~al.}(2019)\citenamefont {Lu},
  \citenamefont {Stepanov}, \citenamefont {Yang}, \citenamefont {Xie},
  \citenamefont {Aamir}, \citenamefont {Das}, \citenamefont {Urgell},
  \citenamefont {Watanabe}, \citenamefont {Taniguchi}, \citenamefont {Zhang},
  \citenamefont {Bachtold}, \citenamefont {MacDonald},\ and\ \citenamefont
  {Efetov}}]{Lu2019}%
  \BibitemOpen
  \bibfield  {author} {\bibinfo {author} {\bibfnamefont {X.}~\bibnamefont
  {Lu}}, \bibinfo {author} {\bibfnamefont {P.}~\bibnamefont {Stepanov}},
  \bibinfo {author} {\bibfnamefont {W.}~\bibnamefont {Yang}}, \bibinfo {author}
  {\bibfnamefont {M.}~\bibnamefont {Xie}}, \bibinfo {author} {\bibfnamefont
  {M.~A.}\ \bibnamefont {Aamir}}, \bibinfo {author} {\bibfnamefont
  {I.}~\bibnamefont {Das}}, \bibinfo {author} {\bibfnamefont {C.}~\bibnamefont
  {Urgell}}, \bibinfo {author} {\bibfnamefont {K.}~\bibnamefont {Watanabe}},
  \bibinfo {author} {\bibfnamefont {T.}~\bibnamefont {Taniguchi}}, \bibinfo
  {author} {\bibfnamefont {G.}~\bibnamefont {Zhang}}, \bibinfo {author}
  {\bibfnamefont {A.}~\bibnamefont {Bachtold}}, \bibinfo {author}
  {\bibfnamefont {A.~H.}\ \bibnamefont {MacDonald}}, \ and\ \bibinfo {author}
  {\bibfnamefont {D.~K.}\ \bibnamefont {Efetov}},\ }\href {\doibase
  10.1038/s41586-019-1695-0} {\bibfield  {journal} {\bibinfo  {journal}
  {Nature}\ }\textbf {\bibinfo {volume} {574}},\ \bibinfo {pages} {653}
  (\bibinfo {year} {2019})}\BibitemShut {NoStop}%
\bibitem [{\citenamefont {Liu}\ \emph {et~al.}(2020)\citenamefont {Liu},
  \citenamefont {Hao}, \citenamefont {Khalaf}, \citenamefont {Lee},
  \citenamefont {Ronen}, \citenamefont {Yoo}, \citenamefont {Najafabadi},
  \citenamefont {Watanabe}, \citenamefont {Taniguchi}, \citenamefont
  {Vishwanath},\ and\ \citenamefont {Kim}}]{Liu2020}%
  \BibitemOpen
  \bibfield  {author} {\bibinfo {author} {\bibfnamefont {X.}~\bibnamefont
  {Liu}}, \bibinfo {author} {\bibfnamefont {Z.}~\bibnamefont {Hao}}, \bibinfo
  {author} {\bibfnamefont {E.}~\bibnamefont {Khalaf}}, \bibinfo {author}
  {\bibfnamefont {J.~Y.}\ \bibnamefont {Lee}}, \bibinfo {author} {\bibfnamefont
  {Y.}~\bibnamefont {Ronen}}, \bibinfo {author} {\bibfnamefont
  {H.}~\bibnamefont {Yoo}}, \bibinfo {author} {\bibfnamefont {D.~H.}\
  \bibnamefont {Najafabadi}}, \bibinfo {author} {\bibfnamefont
  {K.}~\bibnamefont {Watanabe}}, \bibinfo {author} {\bibfnamefont
  {T.}~\bibnamefont {Taniguchi}}, \bibinfo {author} {\bibfnamefont
  {A.}~\bibnamefont {Vishwanath}}, \ and\ \bibinfo {author} {\bibfnamefont
  {P.}~\bibnamefont {Kim}},\ }\href {\doibase 10.1038/s41586-020-2458-7}
  {\bibfield  {journal} {\bibinfo  {journal} {Nature}\ }\textbf {\bibinfo
  {volume} {583}},\ \bibinfo {pages} {221} (\bibinfo {year}
  {2020})}\BibitemShut {NoStop}%
\bibitem [{\citenamefont {Regan}\ \emph {et~al.}(2020)\citenamefont {Regan},
  \citenamefont {Wang}, \citenamefont {Jin}, \citenamefont {Utama},
  \citenamefont {Gao}, \citenamefont {Wei}, \citenamefont {Zhao}, \citenamefont
  {Zhao}, \citenamefont {Zhang}, \citenamefont {Yumigeta}, \citenamefont
  {Blei}, \citenamefont {Carlström}, \citenamefont {Watanabe}, \citenamefont
  {Taniguchi}, \citenamefont {Tongay}, \citenamefont {Crommie}, \citenamefont
  {Zettl},\ and\ \citenamefont {Wang}}]{Regan2020}%
  \BibitemOpen
  \bibfield  {author} {\bibinfo {author} {\bibfnamefont {E.~C.}\ \bibnamefont
  {Regan}}, \bibinfo {author} {\bibfnamefont {D.}~\bibnamefont {Wang}},
  \bibinfo {author} {\bibfnamefont {C.}~\bibnamefont {Jin}}, \bibinfo {author}
  {\bibfnamefont {M.~I.~B.}\ \bibnamefont {Utama}}, \bibinfo {author}
  {\bibfnamefont {B.}~\bibnamefont {Gao}}, \bibinfo {author} {\bibfnamefont
  {X.}~\bibnamefont {Wei}}, \bibinfo {author} {\bibfnamefont {S.}~\bibnamefont
  {Zhao}}, \bibinfo {author} {\bibfnamefont {W.}~\bibnamefont {Zhao}}, \bibinfo
  {author} {\bibfnamefont {Z.}~\bibnamefont {Zhang}}, \bibinfo {author}
  {\bibfnamefont {K.}~\bibnamefont {Yumigeta}}, \bibinfo {author}
  {\bibfnamefont {M.}~\bibnamefont {Blei}}, \bibinfo {author} {\bibfnamefont
  {J.~D.}\ \bibnamefont {Carlström}}, \bibinfo {author} {\bibfnamefont
  {K.}~\bibnamefont {Watanabe}}, \bibinfo {author} {\bibfnamefont
  {T.}~\bibnamefont {Taniguchi}}, \bibinfo {author} {\bibfnamefont
  {S.}~\bibnamefont {Tongay}}, \bibinfo {author} {\bibfnamefont
  {M.}~\bibnamefont {Crommie}}, \bibinfo {author} {\bibfnamefont
  {A.}~\bibnamefont {Zettl}}, \ and\ \bibinfo {author} {\bibfnamefont
  {F.}~\bibnamefont {Wang}},\ }\href {\doibase 10.1038/s41586-020-2092-4}
  {\bibfield  {journal} {\bibinfo  {journal} {Nature}\ }\textbf {\bibinfo
  {volume} {579}},\ \bibinfo {pages} {359} (\bibinfo {year}
  {2020})}\BibitemShut {NoStop}%
\bibitem [{\citenamefont {Tang}\ \emph {et~al.}(2020)\citenamefont {Tang},
  \citenamefont {Li}, \citenamefont {Li}, \citenamefont {Xu}, \citenamefont
  {Liu}, \citenamefont {Barmak}, \citenamefont {Watanabe}, \citenamefont
  {Taniguchi}, \citenamefont {MacDonald}, \citenamefont {Shan},\ and\
  \citenamefont {Mak}}]{Tang2020}%
  \BibitemOpen
  \bibfield  {author} {\bibinfo {author} {\bibfnamefont {Y.}~\bibnamefont
  {Tang}}, \bibinfo {author} {\bibfnamefont {L.}~\bibnamefont {Li}}, \bibinfo
  {author} {\bibfnamefont {T.}~\bibnamefont {Li}}, \bibinfo {author}
  {\bibfnamefont {Y.}~\bibnamefont {Xu}}, \bibinfo {author} {\bibfnamefont
  {S.}~\bibnamefont {Liu}}, \bibinfo {author} {\bibfnamefont {K.}~\bibnamefont
  {Barmak}}, \bibinfo {author} {\bibfnamefont {K.}~\bibnamefont {Watanabe}},
  \bibinfo {author} {\bibfnamefont {T.}~\bibnamefont {Taniguchi}}, \bibinfo
  {author} {\bibfnamefont {A.~H.}\ \bibnamefont {MacDonald}}, \bibinfo {author}
  {\bibfnamefont {J.}~\bibnamefont {Shan}}, \ and\ \bibinfo {author}
  {\bibfnamefont {K.~F.}\ \bibnamefont {Mak}},\ }\href {\doibase
  10.1038/s41586-020-2085-3} {\bibfield  {journal} {\bibinfo  {journal}
  {Nature}\ }\textbf {\bibinfo {volume} {579}},\ \bibinfo {pages} {353}
  (\bibinfo {year} {2020})}\BibitemShut {NoStop}%
\bibitem [{\citenamefont {Shimazaki}\ \emph {et~al.}(2020)\citenamefont
  {Shimazaki}, \citenamefont {Schwartz}, \citenamefont {Watanabe},
  \citenamefont {Taniguchi}, \citenamefont {Kroner},\ and\ \citenamefont
  {Imamo{\u{g}}lu}}]{Shimazaki2020}%
  \BibitemOpen
  \bibfield  {author} {\bibinfo {author} {\bibfnamefont {Y.}~\bibnamefont
  {Shimazaki}}, \bibinfo {author} {\bibfnamefont {I.}~\bibnamefont {Schwartz}},
  \bibinfo {author} {\bibfnamefont {K.}~\bibnamefont {Watanabe}}, \bibinfo
  {author} {\bibfnamefont {T.}~\bibnamefont {Taniguchi}}, \bibinfo {author}
  {\bibfnamefont {M.}~\bibnamefont {Kroner}}, \ and\ \bibinfo {author}
  {\bibfnamefont {A.}~\bibnamefont {Imamo{\u{g}}lu}},\ }\href {\doibase
  10.1038/s41586-020-2191-2} {\bibfield  {journal} {\bibinfo  {journal}
  {Nature}\ }\textbf {\bibinfo {volume} {580}},\ \bibinfo {pages} {472}
  (\bibinfo {year} {2020})}\BibitemShut {NoStop}%
\bibitem [{\citenamefont {Wang}\ \emph {et~al.}(2020)\citenamefont {Wang},
  \citenamefont {Shih}, \citenamefont {Ghiotto}, \citenamefont {Xian},
  \citenamefont {Rhodes}, \citenamefont {Tan}, \citenamefont {Claassen},
  \citenamefont {Kennes}, \citenamefont {Bai}, \citenamefont {Kim},
  \citenamefont {Watanabe}, \citenamefont {Taniguchi}, \citenamefont {Zhu},
  \citenamefont {Hone}, \citenamefont {Rubio}, \citenamefont {Pasupathy},\ and\
  \citenamefont {Dean}}]{Wang2020}%
  \BibitemOpen
  \bibfield  {author} {\bibinfo {author} {\bibfnamefont {L.}~\bibnamefont
  {Wang}}, \bibinfo {author} {\bibfnamefont {E.-M.}\ \bibnamefont {Shih}},
  \bibinfo {author} {\bibfnamefont {A.}~\bibnamefont {Ghiotto}}, \bibinfo
  {author} {\bibfnamefont {L.}~\bibnamefont {Xian}}, \bibinfo {author}
  {\bibfnamefont {D.~A.}\ \bibnamefont {Rhodes}}, \bibinfo {author}
  {\bibfnamefont {C.}~\bibnamefont {Tan}}, \bibinfo {author} {\bibfnamefont
  {M.}~\bibnamefont {Claassen}}, \bibinfo {author} {\bibfnamefont {D.~M.}\
  \bibnamefont {Kennes}}, \bibinfo {author} {\bibfnamefont {Y.}~\bibnamefont
  {Bai}}, \bibinfo {author} {\bibfnamefont {B.}~\bibnamefont {Kim}}, \bibinfo
  {author} {\bibfnamefont {K.}~\bibnamefont {Watanabe}}, \bibinfo {author}
  {\bibfnamefont {T.}~\bibnamefont {Taniguchi}}, \bibinfo {author}
  {\bibfnamefont {X.}~\bibnamefont {Zhu}}, \bibinfo {author} {\bibfnamefont
  {J.}~\bibnamefont {Hone}}, \bibinfo {author} {\bibfnamefont {A.}~\bibnamefont
  {Rubio}}, \bibinfo {author} {\bibfnamefont {A.~N.}\ \bibnamefont
  {Pasupathy}}, \ and\ \bibinfo {author} {\bibfnamefont {C.~R.}\ \bibnamefont
  {Dean}},\ }\href {\doibase 10.1038/s41563-020-0708-6} {\bibfield  {journal}
  {\bibinfo  {journal} {Nature Materials}\ }\textbf {\bibinfo {volume} {19}},\
  \bibinfo {pages} {861} (\bibinfo {year} {2020})}\BibitemShut {NoStop}%
\bibitem [{\citenamefont {Shimazaki}\ \emph {et~al.}(2021)\citenamefont
  {Shimazaki}, \citenamefont {Kuhlenkamp}, \citenamefont {Schwartz},
  \citenamefont {Smole\ifmmode~\acute{n}\else \'{n}\fi{}ski}, \citenamefont
  {Watanabe}, \citenamefont {Taniguchi}, \citenamefont {Kroner}, \citenamefont
  {Schmidt}, \citenamefont {Knap},\ and\ \citenamefont
  {Imamo{\u{g}}lu}}]{Shimazaki2021}%
  \BibitemOpen
  \bibfield  {author} {\bibinfo {author} {\bibfnamefont {Y.}~\bibnamefont
  {Shimazaki}}, \bibinfo {author} {\bibfnamefont {C.}~\bibnamefont
  {Kuhlenkamp}}, \bibinfo {author} {\bibfnamefont {I.}~\bibnamefont
  {Schwartz}}, \bibinfo {author} {\bibfnamefont {T.}~\bibnamefont
  {Smole\ifmmode~\acute{n}\else \'{n}\fi{}ski}}, \bibinfo {author}
  {\bibfnamefont {K.}~\bibnamefont {Watanabe}}, \bibinfo {author}
  {\bibfnamefont {T.}~\bibnamefont {Taniguchi}}, \bibinfo {author}
  {\bibfnamefont {M.}~\bibnamefont {Kroner}}, \bibinfo {author} {\bibfnamefont
  {R.}~\bibnamefont {Schmidt}}, \bibinfo {author} {\bibfnamefont
  {M.}~\bibnamefont {Knap}}, \ and\ \bibinfo {author} {\bibfnamefont
  {A.}~\bibnamefont {Imamo{\u{g}}lu}},\ }\href {\doibase
  10.1103/PhysRevX.11.021027} {\bibfield  {journal} {\bibinfo  {journal} {Phys.
  Rev. X}\ }\textbf {\bibinfo {volume} {11}},\ \bibinfo {pages} {021027}
  (\bibinfo {year} {2021})}\BibitemShut {NoStop}%
\bibitem [{\citenamefont {Sidler}\ \emph {et~al.}(2016)\citenamefont {Sidler},
  \citenamefont {Back}, \citenamefont {Cotlet}, \citenamefont {Srivastava},
  \citenamefont {Fink}, \citenamefont {Kroner}, \citenamefont {Demler},\ and\
  \citenamefont {Imamoglu}}]{Sidler2016}%
  \BibitemOpen
  \bibfield  {author} {\bibinfo {author} {\bibfnamefont {M.}~\bibnamefont
  {Sidler}}, \bibinfo {author} {\bibfnamefont {P.}~\bibnamefont {Back}},
  \bibinfo {author} {\bibfnamefont {O.}~\bibnamefont {Cotlet}}, \bibinfo
  {author} {\bibfnamefont {A.}~\bibnamefont {Srivastava}}, \bibinfo {author}
  {\bibfnamefont {T.}~\bibnamefont {Fink}}, \bibinfo {author} {\bibfnamefont
  {M.}~\bibnamefont {Kroner}}, \bibinfo {author} {\bibfnamefont
  {E.}~\bibnamefont {Demler}}, \ and\ \bibinfo {author} {\bibfnamefont
  {A.}~\bibnamefont {Imamoglu}},\ }\href {\doibase 10.1038/nphys3949}
  {\bibfield  {journal} {\bibinfo  {journal} {Nature Physics}\ }\textbf
  {\bibinfo {volume} {13}},\ \bibinfo {pages} {255} (\bibinfo {year}
  {2016})}\BibitemShut {NoStop}%
\bibitem [{\citenamefont {Efimkin}\ and\ \citenamefont
  {MacDonald}(2017)}]{Efimkin2017}%
  \BibitemOpen
  \bibfield  {author} {\bibinfo {author} {\bibfnamefont {D.~K.}\ \bibnamefont
  {Efimkin}}\ and\ \bibinfo {author} {\bibfnamefont {A.~H.}\ \bibnamefont
  {MacDonald}},\ }\href {\doibase 10.1103/physrevb.95.035417} {\bibfield
  {journal} {\bibinfo  {journal} {Physical Review B}\ }\textbf {\bibinfo
  {volume} {95}},\ \bibinfo {pages} {035417} (\bibinfo {year}
  {2017})}\BibitemShut {NoStop}%
\bibitem [{\citenamefont {Takemura}\ \emph {et~al.}(2014)\citenamefont
  {Takemura}, \citenamefont {Trebaol}, \citenamefont {Wouters}, \citenamefont
  {Portella-Oberli},\ and\ \citenamefont {Deveaud}}]{Takemura2014}%
  \BibitemOpen
  \bibfield  {author} {\bibinfo {author} {\bibfnamefont {N.}~\bibnamefont
  {Takemura}}, \bibinfo {author} {\bibfnamefont {S.}~\bibnamefont {Trebaol}},
  \bibinfo {author} {\bibfnamefont {M.}~\bibnamefont {Wouters}}, \bibinfo
  {author} {\bibfnamefont {M.~T.}\ \bibnamefont {Portella-Oberli}}, \ and\
  \bibinfo {author} {\bibfnamefont {B.}~\bibnamefont {Deveaud}},\ }\href
  {\doibase 10.1038/nphys2999} {\bibfield  {journal} {\bibinfo  {journal}
  {Nature Physics}\ }\textbf {\bibinfo {volume} {10}},\ \bibinfo {pages} {500}
  (\bibinfo {year} {2014})}\BibitemShut {NoStop}%
\bibitem [{\citenamefont {Yu}\ \emph {et~al.}(2019)\citenamefont {Yu},
  \citenamefont {Chen},\ and\ \citenamefont {Yao}}]{Yu2019}%
  \BibitemOpen
  \bibfield  {author} {\bibinfo {author} {\bibfnamefont {H.}~\bibnamefont
  {Yu}}, \bibinfo {author} {\bibfnamefont {M.}~\bibnamefont {Chen}}, \ and\
  \bibinfo {author} {\bibfnamefont {W.}~\bibnamefont {Yao}},\ }\href {\doibase
  10.1093/nsr/nwz117} {\bibfield  {journal} {\bibinfo  {journal} {National
  Science Review}\ }\textbf {\bibinfo {volume} {7}},\ \bibinfo {pages} {12}
  (\bibinfo {year} {2019})}\BibitemShut {NoStop}%
\bibitem [{\citenamefont {Stinaff}\ \emph {et~al.}(2006)\citenamefont
  {Stinaff}, \citenamefont {Scheibner}, \citenamefont {Bracker}, \citenamefont
  {Ponomarev}, \citenamefont {Korenev}, \citenamefont {Ware}, \citenamefont
  {Doty}, \citenamefont {Reinecke},\ and\ \citenamefont
  {Gammon}}]{Stinaff2006}%
  \BibitemOpen
  \bibfield  {author} {\bibinfo {author} {\bibfnamefont {E.~A.}\ \bibnamefont
  {Stinaff}}, \bibinfo {author} {\bibfnamefont {M.}~\bibnamefont {Scheibner}},
  \bibinfo {author} {\bibfnamefont {A.~S.}\ \bibnamefont {Bracker}}, \bibinfo
  {author} {\bibfnamefont {I.~V.}\ \bibnamefont {Ponomarev}}, \bibinfo {author}
  {\bibfnamefont {V.~L.}\ \bibnamefont {Korenev}}, \bibinfo {author}
  {\bibfnamefont {M.~E.}\ \bibnamefont {Ware}}, \bibinfo {author}
  {\bibfnamefont {M.~F.}\ \bibnamefont {Doty}}, \bibinfo {author}
  {\bibfnamefont {T.~L.}\ \bibnamefont {Reinecke}}, \ and\ \bibinfo {author}
  {\bibfnamefont {D.}~\bibnamefont {Gammon}},\ }\href {\doibase
  10.1126/science.1121189} {\bibfield  {journal} {\bibinfo  {journal}
  {Science}\ }\textbf {\bibinfo {volume} {311}},\ \bibinfo {pages} {636}
  (\bibinfo {year} {2006})}\BibitemShut {NoStop}%
\bibitem [{\citenamefont {Krenner}\ \emph {et~al.}(2006)\citenamefont
  {Krenner}, \citenamefont {Clark}, \citenamefont {Nakaoka}, \citenamefont
  {Bichler}, \citenamefont {Scheurer}, \citenamefont {Abstreiter},\ and\
  \citenamefont {Finley}}]{Krenner2006}%
  \BibitemOpen
  \bibfield  {author} {\bibinfo {author} {\bibfnamefont {H.~J.}\ \bibnamefont
  {Krenner}}, \bibinfo {author} {\bibfnamefont {E.~C.}\ \bibnamefont {Clark}},
  \bibinfo {author} {\bibfnamefont {T.}~\bibnamefont {Nakaoka}}, \bibinfo
  {author} {\bibfnamefont {M.}~\bibnamefont {Bichler}}, \bibinfo {author}
  {\bibfnamefont {C.}~\bibnamefont {Scheurer}}, \bibinfo {author}
  {\bibfnamefont {G.}~\bibnamefont {Abstreiter}}, \ and\ \bibinfo {author}
  {\bibfnamefont {J.~J.}\ \bibnamefont {Finley}},\ }\href {\doibase
  10.1103/physrevlett.97.076403} {\bibfield  {journal} {\bibinfo  {journal}
  {Physical Review Letters}\ }\textbf {\bibinfo {volume} {97}},\ \bibinfo
  {pages} {076403} (\bibinfo {year} {2006})}\BibitemShut {NoStop}%
\bibitem [{\citenamefont {Fertig}(1989)}]{Fertig1989}%
  \BibitemOpen
  \bibfield  {author} {\bibinfo {author} {\bibfnamefont {H.~A.}\ \bibnamefont
  {Fertig}},\ }\href {\doibase 10.1103/physrevb.40.1087} {\bibfield  {journal}
  {\bibinfo  {journal} {Physical Review B}\ }\textbf {\bibinfo {volume} {40}},\
  \bibinfo {pages} {1087} (\bibinfo {year} {1989})}\BibitemShut {NoStop}%
\bibitem [{\citenamefont {MacDonald}\ \emph {et~al.}(1990)\citenamefont
  {MacDonald}, \citenamefont {Platzman},\ and\ \citenamefont
  {Boebinger}}]{MacDonald1990}%
  \BibitemOpen
  \bibfield  {author} {\bibinfo {author} {\bibfnamefont {A.~H.}\ \bibnamefont
  {MacDonald}}, \bibinfo {author} {\bibfnamefont {P.~M.}\ \bibnamefont
  {Platzman}}, \ and\ \bibinfo {author} {\bibfnamefont {G.~S.}\ \bibnamefont
  {Boebinger}},\ }\href {\doibase 10.1103/physrevlett.65.775} {\bibfield
  {journal} {\bibinfo  {journal} {Physical Review Letters}\ }\textbf {\bibinfo
  {volume} {65}},\ \bibinfo {pages} {775} (\bibinfo {year} {1990})}\BibitemShut
  {NoStop}%
\bibitem [{\citenamefont {Kohstall}\ \emph {et~al.}(2012)\citenamefont
  {Kohstall}, \citenamefont {Zaccanti}, \citenamefont {Jag}, \citenamefont
  {Trenkwalder}, \citenamefont {Massignan}, \citenamefont {Bruun},
  \citenamefont {Schreck},\ and\ \citenamefont {Grimm}}]{Kohstall2012}%
  \BibitemOpen
  \bibfield  {author} {\bibinfo {author} {\bibfnamefont {C.}~\bibnamefont
  {Kohstall}}, \bibinfo {author} {\bibfnamefont {M.}~\bibnamefont {Zaccanti}},
  \bibinfo {author} {\bibfnamefont {M.}~\bibnamefont {Jag}}, \bibinfo {author}
  {\bibfnamefont {A.}~\bibnamefont {Trenkwalder}}, \bibinfo {author}
  {\bibfnamefont {P.}~\bibnamefont {Massignan}}, \bibinfo {author}
  {\bibfnamefont {G.~M.}\ \bibnamefont {Bruun}}, \bibinfo {author}
  {\bibfnamefont {F.}~\bibnamefont {Schreck}}, \ and\ \bibinfo {author}
  {\bibfnamefont {R.}~\bibnamefont {Grimm}},\ }\href {\doibase
  10.1038/nature11065} {\bibfield  {journal} {\bibinfo  {journal} {Nature}\
  }\textbf {\bibinfo {volume} {485}},\ \bibinfo {pages} {615} (\bibinfo {year}
  {2012})}\BibitemShut {NoStop}%
\bibitem [{\citenamefont {Schmidt}\ \emph {et~al.}(2012)\citenamefont
  {Schmidt}, \citenamefont {Enss}, \citenamefont {Pietil\"a},\ and\
  \citenamefont {Demler}}]{Schmidt2012}%
  \BibitemOpen
  \bibfield  {author} {\bibinfo {author} {\bibfnamefont {R.}~\bibnamefont
  {Schmidt}}, \bibinfo {author} {\bibfnamefont {T.}~\bibnamefont {Enss}},
  \bibinfo {author} {\bibfnamefont {V.}~\bibnamefont {Pietil\"a}}, \ and\
  \bibinfo {author} {\bibfnamefont {E.}~\bibnamefont {Demler}},\ }\href
  {\doibase 10.1103/PhysRevA.85.021602} {\bibfield  {journal} {\bibinfo
  {journal} {Phys. Rev. A}\ }\textbf {\bibinfo {volume} {85}},\ \bibinfo
  {pages} {021602} (\bibinfo {year} {2012})}\BibitemShut {NoStop}%
\bibitem [{\citenamefont {Burch}\ \emph {et~al.}(2018)\citenamefont {Burch},
  \citenamefont {Mandrus},\ and\ \citenamefont {Park}}]{Burch2018}%
  \BibitemOpen
  \bibfield  {author} {\bibinfo {author} {\bibfnamefont {K.~S.}\ \bibnamefont
  {Burch}}, \bibinfo {author} {\bibfnamefont {D.}~\bibnamefont {Mandrus}}, \
  and\ \bibinfo {author} {\bibfnamefont {J.-G.}\ \bibnamefont {Park}},\ }\href
  {\doibase 10.1038/s41586-018-0631-z} {\bibfield  {journal} {\bibinfo
  {journal} {Nature}\ }\textbf {\bibinfo {volume} {563}},\ \bibinfo {pages}
  {47} (\bibinfo {year} {2018})}\BibitemShut {NoStop}%
\bibitem [{\citenamefont {Slagle}\ and\ \citenamefont {Fu}(2020)}]{Slagle2020}%
  \BibitemOpen
  \bibfield  {author} {\bibinfo {author} {\bibfnamefont {K.}~\bibnamefont
  {Slagle}}\ and\ \bibinfo {author} {\bibfnamefont {L.}~\bibnamefont {Fu}},\
  }\href {\doibase 10.1103/physrevb.102.235423} {\bibfield  {journal} {\bibinfo
   {journal} {Physical Review B}\ }\textbf {\bibinfo {volume} {102}},\ \bibinfo
  {pages} {235423} (\bibinfo {year} {2020})}\BibitemShut {NoStop}%
\bibitem [{\citenamefont {Zhang}\ \emph {et~al.}(2021)\citenamefont {Zhang},
  \citenamefont {Liu},\ and\ \citenamefont {Fu}}]{Zhang2021}%
  \BibitemOpen
  \bibfield  {author} {\bibinfo {author} {\bibfnamefont {Y.}~\bibnamefont
  {Zhang}}, \bibinfo {author} {\bibfnamefont {T.}~\bibnamefont {Liu}}, \ and\
  \bibinfo {author} {\bibfnamefont {L.}~\bibnamefont {Fu}},\ }\href {\doibase
  10.1103/physrevb.103.155142} {\bibfield  {journal} {\bibinfo  {journal}
  {Physical Review B}\ }\textbf {\bibinfo {volume} {103}},\ \bibinfo {pages}
  {155142} (\bibinfo {year} {2021})}\BibitemShut {NoStop}%
\end{thebibliography}%


\begin{thebibliography}{5}%
\makeatletter
\providecommand \@ifxundefined [1]{%
 \@ifx{#1\undefined}
}%
\providecommand \@ifnum [1]{%
 \ifnum #1\expandafter \@firstoftwo
 \else \expandafter \@secondoftwo
 \fi
}%
\providecommand \@ifx [1]{%
 \ifx #1\expandafter \@firstoftwo
 \else \expandafter \@secondoftwo
 \fi
}%
\providecommand \natexlab [1]{#1}%
\providecommand \enquote  [1]{``#1''}%
\providecommand \bibnamefont  [1]{#1}%
\providecommand \bibfnamefont [1]{#1}%
\providecommand \citenamefont [1]{#1}%
\providecommand \href@noop [0]{\@secondoftwo}%
\providecommand \href [0]{\begingroup \@sanitize@url \@href}%
\providecommand \@href[1]{\@@startlink{#1}\@@href}%
\providecommand \@@href[1]{\endgroup#1\@@endlink}%
\providecommand \@sanitize@url [0]{\catcode `\\12\catcode `\$12\catcode
  `\&12\catcode `\#12\catcode `\^12\catcode `\_12\catcode `\%12\relax}%
\providecommand \@@startlink[1]{}%
\providecommand \@@endlink[0]{}%
\providecommand \url  [0]{\begingroup\@sanitize@url \@url }%
\providecommand \@url [1]{\endgroup\@href {#1}{\urlprefix }}%
\providecommand \urlprefix  [0]{URL }%
\providecommand \Eprint [0]{\href }%
\providecommand \doibase [0]{http://dx.doi.org/}%
\providecommand \selectlanguage [0]{\@gobble}%
\providecommand \bibinfo  [0]{\@secondoftwo}%
\providecommand \bibfield  [0]{\@secondoftwo}%
\providecommand \translation [1]{[#1]}%
\providecommand \BibitemOpen [0]{}%
\providecommand \bibitemStop [0]{}%
\providecommand \bibitemNoStop [0]{.\EOS\space}%
\providecommand \EOS [0]{\spacefactor3000\relax}%
\providecommand \BibitemShut  [1]{\csname bibitem#1\endcsname}%
\let\auto@bib@innerbib\@empty
\bibitem [{\citenamefont {Wang}\ \emph {et~al.}(2013)\citenamefont {Wang},
  \citenamefont {Meric}, \citenamefont {Huang}, \citenamefont {Gao},
  \citenamefont {Gao}, \citenamefont {Tran}, \citenamefont {Taniguchi},
  \citenamefont {Watanabe}, \citenamefont {Campos}, \citenamefont {Muller},
  \citenamefont {Guo}, \citenamefont {Kim}, \citenamefont {Hone}, \citenamefont
  {Shepard},\ and\ \citenamefont {Dean}}]{Wang2013}%
  \BibitemOpen
  \bibfield  {author} {\bibinfo {author} {\bibfnamefont {L.}~\bibnamefont
  {Wang}}, \bibinfo {author} {\bibfnamefont {I.}~\bibnamefont {Meric}},
  \bibinfo {author} {\bibfnamefont {P.~Y.}\ \bibnamefont {Huang}}, \bibinfo
  {author} {\bibfnamefont {Q.}~\bibnamefont {Gao}}, \bibinfo {author}
  {\bibfnamefont {Y.}~\bibnamefont {Gao}}, \bibinfo {author} {\bibfnamefont
  {H.}~\bibnamefont {Tran}}, \bibinfo {author} {\bibfnamefont {T.}~\bibnamefont
  {Taniguchi}}, \bibinfo {author} {\bibfnamefont {K.}~\bibnamefont {Watanabe}},
  \bibinfo {author} {\bibfnamefont {L.~M.}\ \bibnamefont {Campos}}, \bibinfo
  {author} {\bibfnamefont {D.~A.}\ \bibnamefont {Muller}}, \bibinfo {author}
  {\bibfnamefont {J.}~\bibnamefont {Guo}}, \bibinfo {author} {\bibfnamefont
  {P.}~\bibnamefont {Kim}}, \bibinfo {author} {\bibfnamefont {J.}~\bibnamefont
  {Hone}}, \bibinfo {author} {\bibfnamefont {K.~L.}\ \bibnamefont {Shepard}}, \
  and\ \bibinfo {author} {\bibfnamefont {C.~R.}\ \bibnamefont {Dean}},\ }\href
  {\doibase 10.1126/science.1244358} {\bibfield  {journal} {\bibinfo  {journal}
  {Science}\ }\textbf {\bibinfo {volume} {342}},\ \bibinfo {pages} {614}
  (\bibinfo {year} {2013})}\BibitemShut {NoStop}%
\bibitem [{\citenamefont {Kim}\ \emph {et~al.}(2016)\citenamefont {Kim},
  \citenamefont {Yankowitz}, \citenamefont {Fallahazad}, \citenamefont {Kang},
  \citenamefont {Movva}, \citenamefont {Huang}, \citenamefont {Larentis},
  \citenamefont {Corbet}, \citenamefont {Taniguchi}, \citenamefont {Watanabe},
  \citenamefont {Banerjee}, \citenamefont {LeRoy},\ and\ \citenamefont
  {Tutuc}}]{Kim2016}%
  \BibitemOpen
  \bibfield  {author} {\bibinfo {author} {\bibfnamefont {K.}~\bibnamefont
  {Kim}}, \bibinfo {author} {\bibfnamefont {M.}~\bibnamefont {Yankowitz}},
  \bibinfo {author} {\bibfnamefont {B.}~\bibnamefont {Fallahazad}}, \bibinfo
  {author} {\bibfnamefont {S.}~\bibnamefont {Kang}}, \bibinfo {author}
  {\bibfnamefont {H.~C.~P.}\ \bibnamefont {Movva}}, \bibinfo {author}
  {\bibfnamefont {S.}~\bibnamefont {Huang}}, \bibinfo {author} {\bibfnamefont
  {S.}~\bibnamefont {Larentis}}, \bibinfo {author} {\bibfnamefont {C.~M.}\
  \bibnamefont {Corbet}}, \bibinfo {author} {\bibfnamefont {T.}~\bibnamefont
  {Taniguchi}}, \bibinfo {author} {\bibfnamefont {K.}~\bibnamefont {Watanabe}},
  \bibinfo {author} {\bibfnamefont {S.~K.}\ \bibnamefont {Banerjee}}, \bibinfo
  {author} {\bibfnamefont {B.~J.}\ \bibnamefont {LeRoy}}, \ and\ \bibinfo
  {author} {\bibfnamefont {E.}~\bibnamefont {Tutuc}},\ }\href {\doibase
  10.1021/acs.nanolett.5b05263} {\bibfield  {journal} {\bibinfo  {journal}
  {Nano Letters}\ }\textbf {\bibinfo {volume} {16}},\ \bibinfo {pages} {1989}
  (\bibinfo {year} {2016})}\BibitemShut {NoStop}%
\bibitem [{\citenamefont {Shimazaki}\ \emph {et~al.}(2020)\citenamefont
  {Shimazaki}, \citenamefont {Schwartz}, \citenamefont {Watanabe},
  \citenamefont {Taniguchi}, \citenamefont {Kroner},\ and\ \citenamefont
  {Imamo{\u{g}}lu}}]{Shimazaki2020}%
  \BibitemOpen
  \bibfield  {author} {\bibinfo {author} {\bibfnamefont {Y.}~\bibnamefont
  {Shimazaki}}, \bibinfo {author} {\bibfnamefont {I.}~\bibnamefont {Schwartz}},
  \bibinfo {author} {\bibfnamefont {K.}~\bibnamefont {Watanabe}}, \bibinfo
  {author} {\bibfnamefont {T.}~\bibnamefont {Taniguchi}}, \bibinfo {author}
  {\bibfnamefont {M.}~\bibnamefont {Kroner}}, \ and\ \bibinfo {author}
  {\bibfnamefont {A.}~\bibnamefont {Imamo{\u{g}}lu}},\ }\href {\doibase
  10.1038/s41586-020-2191-2} {\bibfield  {journal} {\bibinfo  {journal}
  {Nature}\ }\textbf {\bibinfo {volume} {580}},\ \bibinfo {pages} {472}
  (\bibinfo {year} {2020})}\BibitemShut {NoStop}%
\bibitem [{\citenamefont {Shimazaki}\ \emph {et~al.}(2021)\citenamefont
  {Shimazaki}, \citenamefont {Kuhlenkamp}, \citenamefont {Schwartz},
  \citenamefont {Smole\ifmmode~\acute{n}\else \'{n}\fi{}ski}, \citenamefont
  {Watanabe}, \citenamefont {Taniguchi}, \citenamefont {Kroner}, \citenamefont
  {Schmidt}, \citenamefont {Knap},\ and\ \citenamefont
  {Imamo{\u{g}}lu}}]{Shimazaki2021}%
  \BibitemOpen
  \bibfield  {author} {\bibinfo {author} {\bibfnamefont {Y.}~\bibnamefont
  {Shimazaki}}, \bibinfo {author} {\bibfnamefont {C.}~\bibnamefont
  {Kuhlenkamp}}, \bibinfo {author} {\bibfnamefont {I.}~\bibnamefont
  {Schwartz}}, \bibinfo {author} {\bibfnamefont {T.}~\bibnamefont
  {Smole\ifmmode~\acute{n}\else \'{n}\fi{}ski}}, \bibinfo {author}
  {\bibfnamefont {K.}~\bibnamefont {Watanabe}}, \bibinfo {author}
  {\bibfnamefont {T.}~\bibnamefont {Taniguchi}}, \bibinfo {author}
  {\bibfnamefont {M.}~\bibnamefont {Kroner}}, \bibinfo {author} {\bibfnamefont
  {R.}~\bibnamefont {Schmidt}}, \bibinfo {author} {\bibfnamefont
  {M.}~\bibnamefont {Knap}}, \ and\ \bibinfo {author} {\bibfnamefont
  {A.}~\bibnamefont {Imamo{\u{g}}lu}},\ }\href {\doibase
  10.1103/PhysRevX.11.021027} {\bibfield  {journal} {\bibinfo  {journal} {Phys.
  Rev. X}\ }\textbf {\bibinfo {volume} {11}},\ \bibinfo {pages} {021027}
  (\bibinfo {year} {2021})}\BibitemShut {NoStop}%
\bibitem [{\citenamefont {Yu}\ \emph {et~al.}(2019)\citenamefont {Yu},
  \citenamefont {Chen},\ and\ \citenamefont {Yao}}]{Yu2019}%
  \BibitemOpen
  \bibfield  {author} {\bibinfo {author} {\bibfnamefont {H.}~\bibnamefont
  {Yu}}, \bibinfo {author} {\bibfnamefont {M.}~\bibnamefont {Chen}}, \ and\
  \bibinfo {author} {\bibfnamefont {W.}~\bibnamefont {Yao}},\ }\href {\doibase
  10.1093/nsr/nwz117} {\bibfield  {journal} {\bibinfo  {journal} {National
  Science Review}\ }\textbf {\bibinfo {volume} {7}},\ \bibinfo {pages} {12}
  (\bibinfo {year} {2019})}\BibitemShut {NoStop}%
\end{thebibliography}%

\end{document}


\begin{center}
{\Large \bf Supplementary Materials}
\end{center}
\tableofcontents

\newpage

\section{M\lowercase{aterials and} M\lowercase{ethods}} \label{S1}


\subsection{Device structure} \label{S11}
Figure 1A in the main text shows a schematic of the device that we use in our experiment. The device consists of two \MoSe\ monolayers encapsulated between two thick hBN layers and separated by a single layer hBN tunnel barrier. Top and bottom gate voltages are applied using transparent thin graphene sheets (few layers) at the top and bottom of the device, while the two TMD layers are grounded. All the flakes that we use in the device were mechanically exfoliated and were assembled using the dry transfer technique.~\cite{Wang2013}
We used the tear and stack technique to obtain close to $0^\circ$ alignment between the two \MoSe\ monolayers.~\cite{Kim2016}
In a previous work based on the same device we estimated the twist angle to be $0.8^\circ$, which corresponds to a \moire\ lattice constant of $a_{\rm M} \sim 25~{\rm nm}$ (for more details see Ref.~\cite{Shimazaki2020}).

\subsection{Optical spectroscopy} \label{S12}
All the PL and differential reflection measurements were performed at cryogenic temperatures of $T\approx 4\,\rm K$. We used a HeNe laser ($633\,\rm nm$) as the PL excitation source and a  fiber-coupled light emitting diode centered at $760\,\rm nm$ with a bandwidth of $20\,\rm nm$ for differential reflection spectroscopy. In all measurements we focused the light using a long working distance apochromatic objective lens with numerical aperture of $0.65$ (attocube LT-APO/LWD/VISIR/0.65). The emitted/reflected light was analyzed using a $0.75\,\rm m$ spectrometer and recorded on a CCD camera that was mounted in the exit port of the spectrometer.
All the reflection data are presented using rolling with window of $0.12\,\rm meV$ for better clarity.

\subsection{Electrical control of the device} \label{S13}
We set \VE=0.5\Vtg-0.5\Vbg\ and \Vmu=0.528\Vtg+0.472\Vbg\ that results in a minimal change to the chemical potential $\mu$ and the electric field $E_z$, respectively.
We note that in the low hole density regime we do not observe a linear $V_\mu$ dependence of carrier filling, probably due to nearby defects or bad contacts, and consequently, the exact density cannot be estimated with a capacitance model. Instead, we estimate the hole density through the emergence of umklapp resonances at integer filling~\cite{Shimazaki2021}.
We further note that at very low filling factors ($\nu \ll 1$), the exact hole density depends both on the intensity of the excitation light and on the applied electric field.

\subsection{U\lowercase{mklapp resonances at integer filling}} \label{S14}
\begin{figure*}[h!]
\centering
\includegraphics[width=1\textwidth]{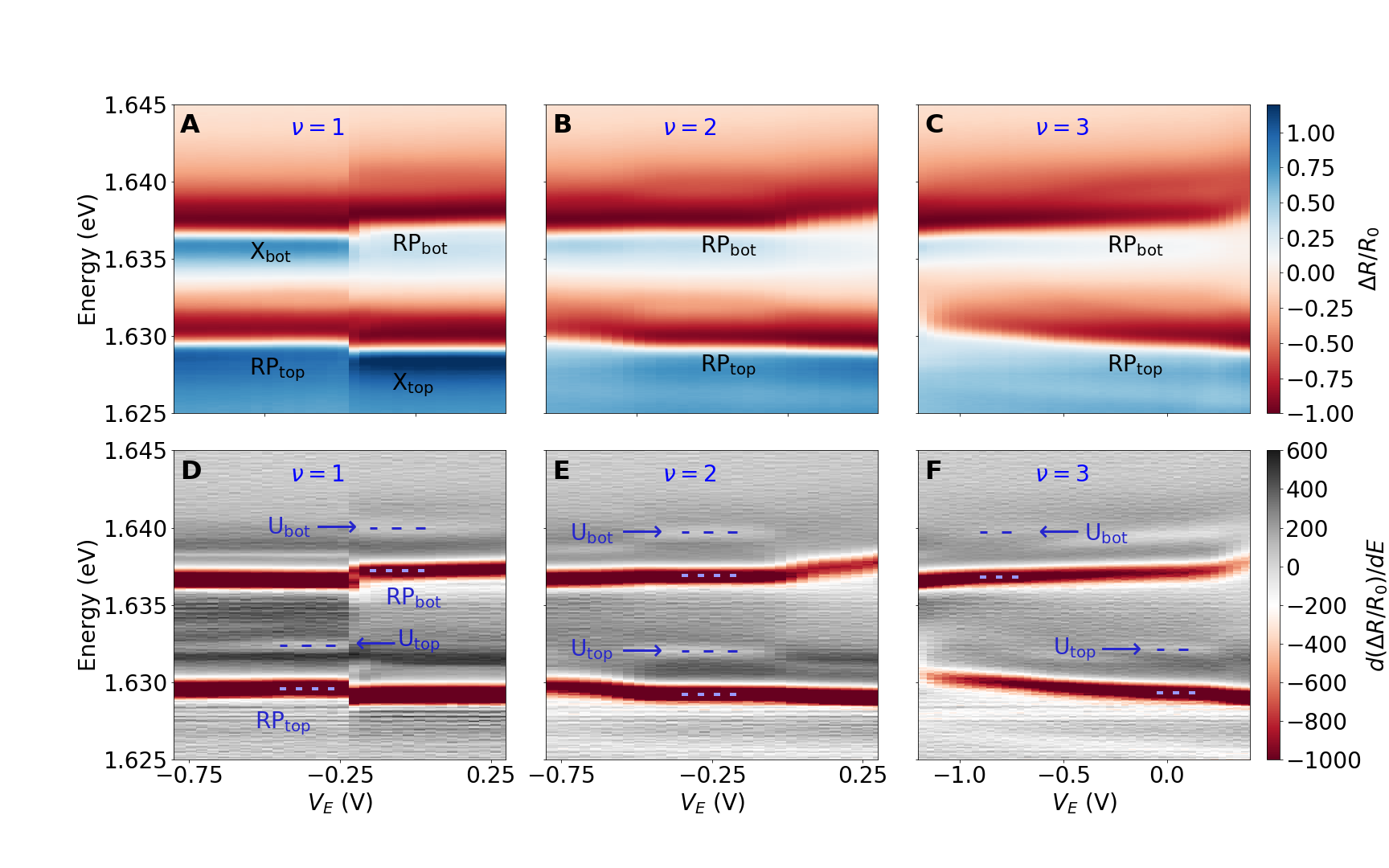}
\caption{{\bf Umklapp resonances at integer fillings.}
        {({\bf A-C})} $V_{\rm E}$ dependent differential reflectance spectra at a fixed chemical potential for $\nu=1$ ({\bf A}), $\nu=2$ ({\bf B}) and $\nu=3$ ({\bf C}). Top (bottom) layer exciton is labeled $\text{X}_{\text{top}}$ ($\text{X}_{\text{bot}}$), top (bottom) layer repulsive polaron is labeled $\text{RP}_{\text{top}}$ ($\text{RP}_{\text{bot}}$).
        {({\bf D-F})} $V_{\rm E}$ dependent differential reflectance spectra differentiated with respect to energy E at a fixed chemical potential for $\nu=1$ ({\bf D}), $\nu=2$ ({\bf E}) and $\nu=3$ ({\bf F}). Top (bottom) layer umklapp resonance is labeled $\rm U_{top}$ ($\rm U_{bot}$). Blue dashed lines marks the RPs and the umklapp scattered resonances.}
\label{figS1}
\end{figure*}

\begin{figure*}[h!]
\centering
\includegraphics[width=1\textwidth]{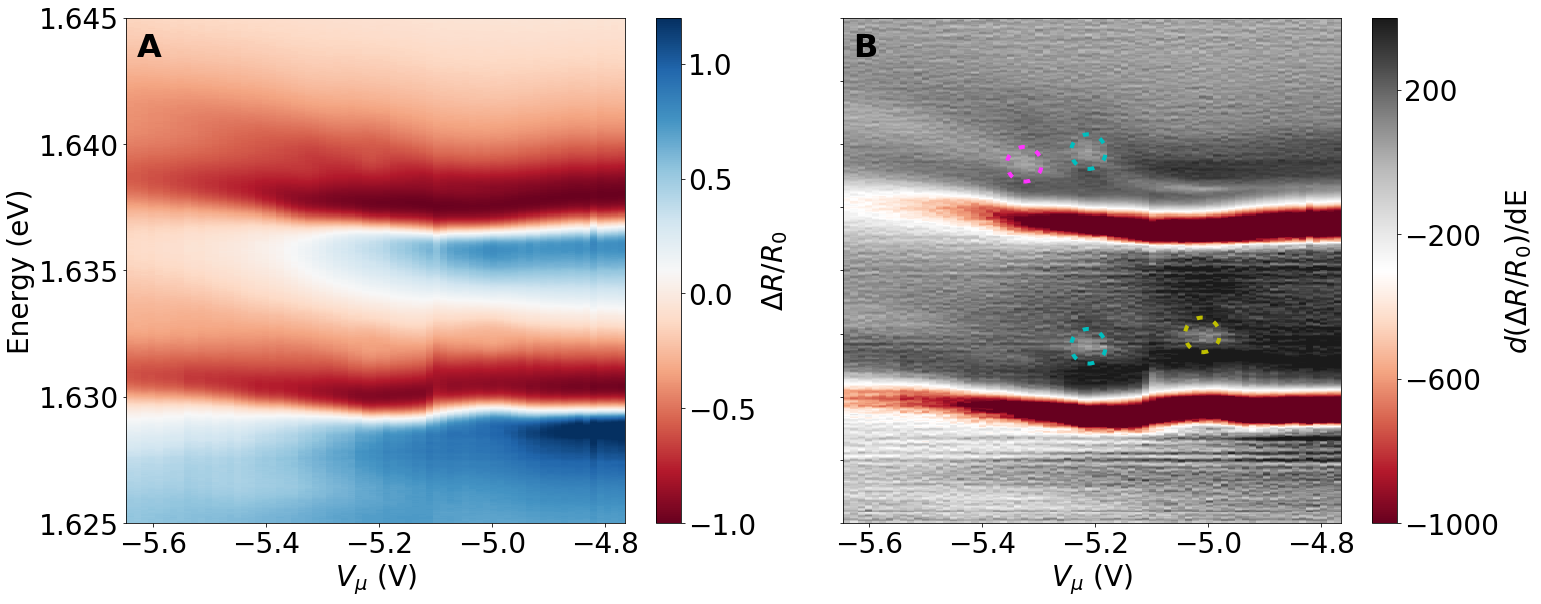}
\caption{{\bf Filling dependence of umklapp resonances.}
        {({\bf A})} $V_{\rm \mu}$ dependent differential reflectance spectra at a fixed electrical field ($V_E=-0.309\,\rm V$). The higher energy resonance is the bottom (top) layer repulsive polaron.
        {({\bf B})} Energy derivative of $V_{\rm \mu}$ dependent differential reflectance spectra at a fixed electrical field ($V_E=-0.309\,\rm V$). The higher (lower) energy resonance is the bottom (top) layer repulsive polaron.
        Yellow dashed circle marks the top layer umklapp resonance at $(\nu_{top},\nu_{bot})=(1,0)$.
        Cyan dashed circles are marking the top and the bottom layers umklapp resonances at $(\nu_{top},\nu_{bot})=(1,1)$.
        Magenta dashed circle marks the bottom layer umklapp resonance at $(\nu_{top},\nu_{bot})\sim(2,1)$.
} \label{figS2}
\end{figure*}

We use the emergence of umklapp resonances in order to evaluate integer filling of the \moire\ subbands, where an incompressible state of holes occurs.~\cite{Shimazaki2020}
Figures~\ref{figS1}A-S1C show \VE\ dependent differential reflection spectra at a fixed chemical potential ($V_\mu=-4.985\,\rm V$, $V_\mu=-5.212\,\rm V$, $V_\mu=-5.325\,\rm V$) that corresponds to $\nu=1$, $\nu=2$ and $\nu=3$, respectively.
Figures~\ref{figS1}D-S1F show the derivative of the \VE\ dependent differential reflection spectra with respect to energy in Figs.~\ref{figS1}A-S1C, respectively.
The top (bottom) layer umklapp resonance appears $\sim 2.8\,\rm meV$ above $\rm RP_{top}$ ($\rm RP_{bot}$) and is labeled $\rm U_{top}$ ($\rm U_{bot}$).
Figure~\ref{figS2}A shows \Vmu\ dependent differential reflectance spectra at a fixed electric field ($V_{\rm E}=-0.309\,\rm V$), where the top layer umklapp resonance at $\nu=1$ ($(\nu_{top},\nu_{bot})=(1,0)$) is marked by a yellow circle.
At a chemical potential corresponding to $\nu=2$ ($(\nu_{top},\nu_{bot})=(1,1)$) the top and bottom umklapp resonances are marked with cyan circles. At a chemical potential corresponding to $\nu=3$ ($(\nu_{top},\nu_{bot})\sim(2,1)$) the bottom layer umklapp resonance is marked with magenta circle.

\section{D\lowercase{escription of the model used to calculate the spectrum}} \label{S2}
We use a non-interacting hole model to describe the order of filling of the \moire\ subbands. As we ignore interactions between the holes, our model cannot describe the rich many body physics that emerges from the interplay between the \moire\ potential and the strong Coulomb interactions.
The \moire\ unit cell exhibit three high symmetry displacement sites between the top and bottom layers atomic registry: these are $\rm MM$, $\rm MX$ and $\rm XM$, where  in $\rm MM$ site the metal atoms of the two layers are aligned and for $\rm MX$ ($\rm XM$) the top (bottom) layer metal atom is aligned with the bottom (top) layer chalcogen atom.
We note that the hole wavefunction at the $\rm MM$ site exhibits inter-layer tunnel coupling, while at $\rm XM$ ($\rm MX$) site there is no inter-layer coupling.~\cite{Yu2019}

The layout of the model is set using our experimental observations. First and foremost, we use the absence of avoided crossing signatures below $\nu\le 1$ (Fig. 1 in the main text) and its appearance for $\nu> 1$ (Fig. 3 in the main text) to conclude that the $\rm XM^t$ ($\rm MX^b$) is the lowest energy site in the top (bottom) layer, in agreement with previous predictions~\cite{Yu2019}. Therefore, for $E_z > 0$ ($E_z < 0$), the injected holes occupy the top (bottom) layer until each $\rm XM^t$ ($\rm XM^b$) site is occupied ($\nu = 1$). The sharp transfer of holes from the top to bottom layer at $E_z =0$ for $\nu = 1$ indicates a first order phase transition and further confirms that the lowest energy sites in the top ($\rm XM^t$) and bottom ($\rm XM^b$) layers are spatially displaced. The appearance of avoided crossing for $\nu > 1$ in turn demonstrates that the second lowest energy site for both layers is the $\rm MM$ site.

To describe the dominant spectral features in \DR\ and photoluminescence (PL) measurements for $E_z \sim 0$, we use a single-particle Hamiltonians describing the ground state and the optically excited state at the $\rm  MM^t,~XM^t,~MX^t,~MM^b,~MX^b,~XM^b $ sites of the moire lattice, with t and b denoting the top and bottom hole states. The initial (ground) state of the optical transition in a \DR\ experiment is described by 
\begin{equation}
    H_g= 
    \begin{pmatrix}
        \Delta/2 & 0                      & 0                       & -t        & 0 & 0 \\
        0        & \Delta/2 + \delta_{XM} & 0                       & 0         & 0 & 0 \\
        0        & 0                      & \Delta/2  + \delta_{MX} & 0         & 0 & 0 \\
        -t       & 0                      & 0                       & -\Delta/2 & 0 & 0 \\          
        0        & 0                      & 0                       & 0         & -\Delta/2 + \delta_{MX} & 0 \\
        0        & 0                      & 0                       & 0         & 0 & -\Delta/2  + \delta_{XM}
\end{pmatrix}
\end{equation}
The optically excited state manifold around the AP resonance in turn is described by
\begin{equation}
H_e= 
\begin{pmatrix}
   \Delta/2 & 0                        & 0                        & 0         & 0                     & 0 \\
    0            & \Delta/2+\delta_{XM} & 0                        & 0         & 0                     & 0 \\
   0            & 0                        & \Delta/2+\delta_{MX} & 0         & 0                     & 0 \\
   0            & 0                        & 0                        & -\Delta/2 + \delta_{bt} & 0                     & 0 \\   
    0            & 0                        & 0                        & 0         & -\Delta/2+\delta_{MX}+ \delta_{bt} & 0 \\
    0            & 0                        & 0                        & 0         & 0                     & -\Delta/2 +\delta_{XM}+ \delta_{bt}
\end{pmatrix}
 + (E_{X}^t - U_t) \cdot \mathbb{I} \rm ,
\end{equation}
with $\mathbb{I}$ the identity matrix and $E_{X}^t$ ($1.6295\,\rm eV$) is the top layer exciton energy and $\delta_{bt}$ ($7.5\,\rm meV$) is the strain induced difference in exciton energies of the bottom and top layers. $U_t$ ($25$~meV) denotes the trion binding energy which is identical for both layers. Both Hamiltonians are written in the site basis $\rm \{ MM^t,~XM^t,~MX^t,~MM^b,~MX^b,~XM^b \}$. The optical transition between these basis states of the initial and excited state Hamiltonians is local, implying that an initial state hole at $\rm XM^t$ site can only create an AP at the same site; the coupling strength however, can be assumed to be identical for all sites. We remark that the hole states in the ground-state manifold corresponding to $\rm MM^t$ and $\rm MM^b$ sites hybridize due to coherent inter-layer tunneling $t$ to form layer-hybridized symmetric ($\rm MM^-$) and antisymmetric ($\rm MM^+$) states. In contrast, large $U_t$ ensures that inter-layer coupling for the excited states can be neglected for $E_z \sim 0$.

\begin{figure*}[h]
\centering
\includegraphics[width=1\textwidth]{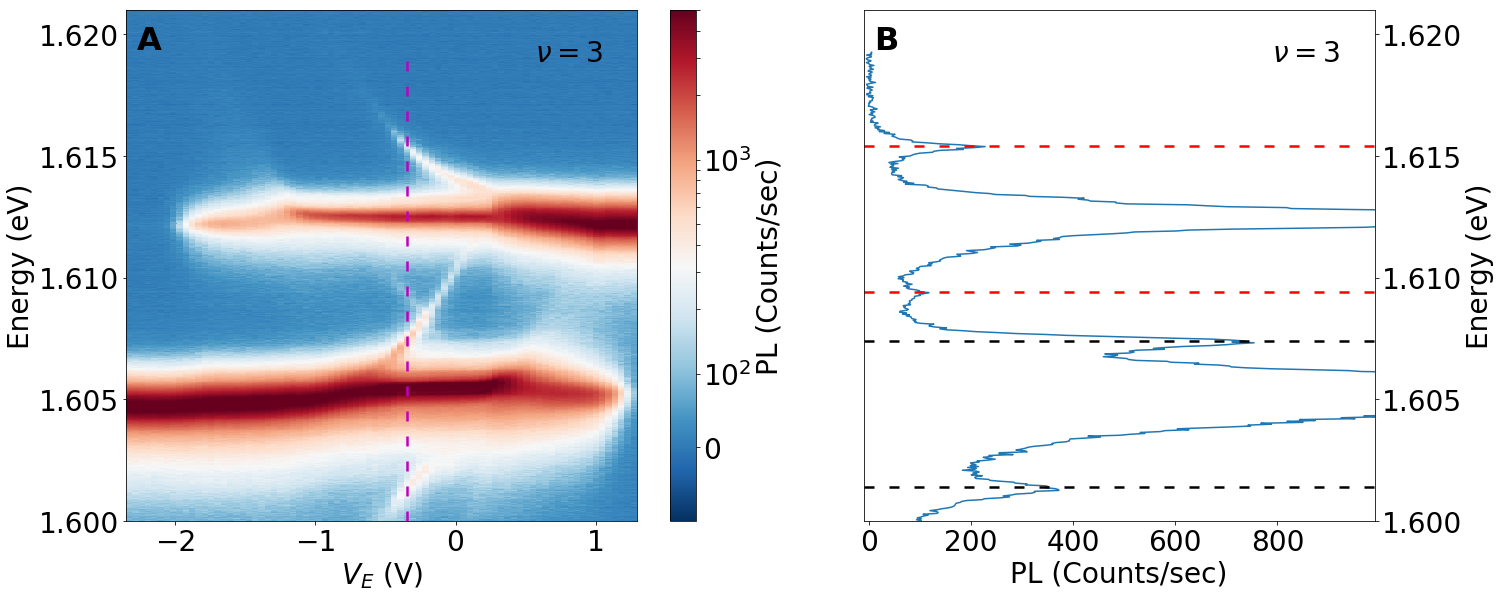}
\caption{{\bf Coupling strength.}
        {({\bf A})} $V_{\rm E}$ dependent PL spectra at a fixed chemical potential for $\nu=3$.
        {({\bf B})} PL linecut along the magenta line in Fig.~\ref{figS3}({\bf A}) ($V_E=-0.345\,\rm V$). The red (black) dashed lines marks the energy of the bottom (top) layer APs that originates from the hybridized hole state.
} \label{figS3}
\end{figure*}

\begin{figure*}[h!]
\centering
\includegraphics[width=1\textwidth]{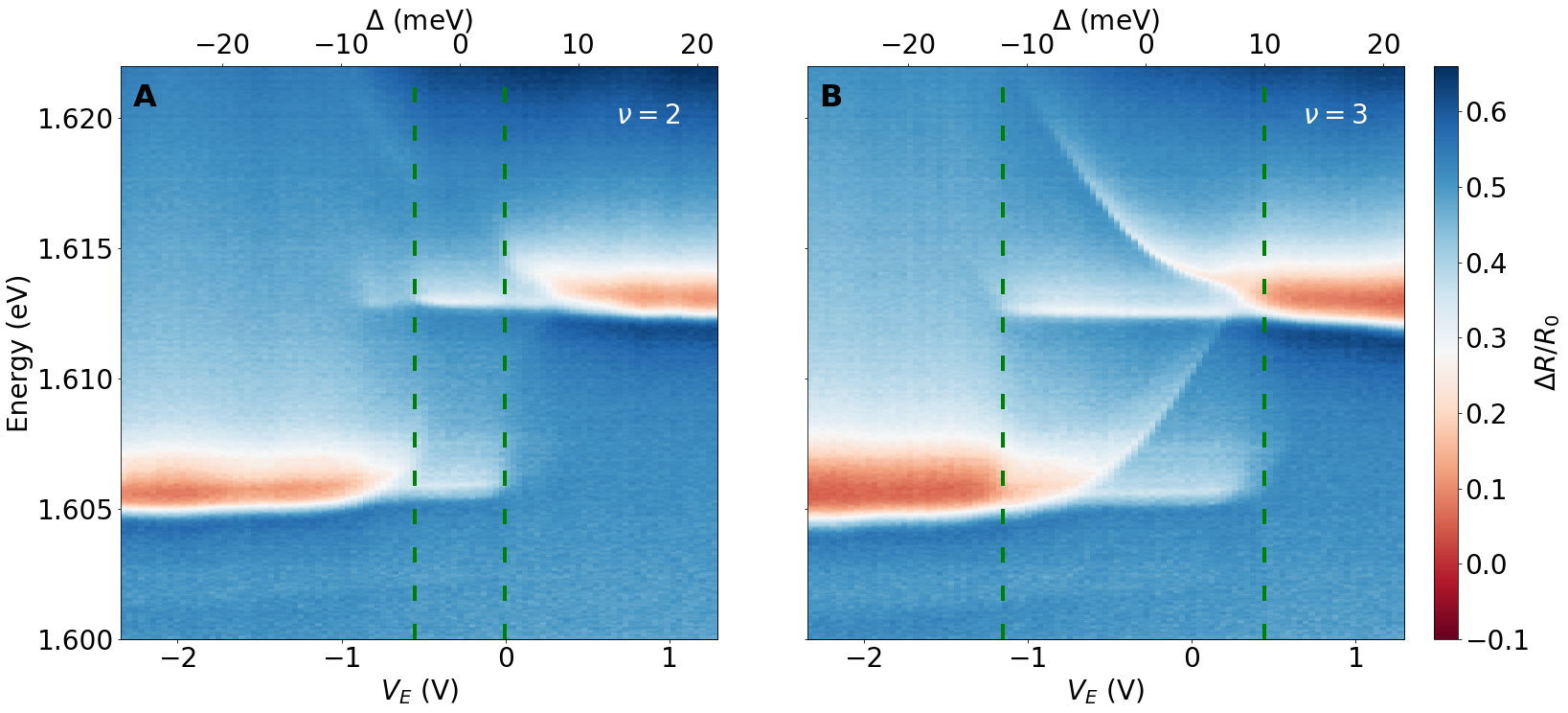}
\caption{{\bf Conversion of \VE\ to energy detuning.}
        {({\bf A})} $V_{\rm E}$ (bottom x axis) and energy detuning (top x axis) dependent differential reflectance spectra at $\nu=2$. Green dashed lines are marking the gap in the avoided crossing.
        {({\bf B})} $V_{\rm E}$ (bottom x axis) and energy detuning (top x axis) dependent differential reflectance spectra at $\nu=3$.
        Green dashed lines are marking the transition of holes from ($\nu_{\rm XM^t}$,$\nu_{\rm MX^b}$)=(1,1) to ($\nu_{\rm XM^t}$,$\nu_{\rm MX^b}$)=(2,0) (left) and ($\nu_{\rm XM^t}$,$\nu_{\rm MX^b}$)=(0,2) (right).
} \label{figS4}
\end{figure*}

\begin{figure*}[h!]
\centering
\includegraphics[width=1\textwidth]{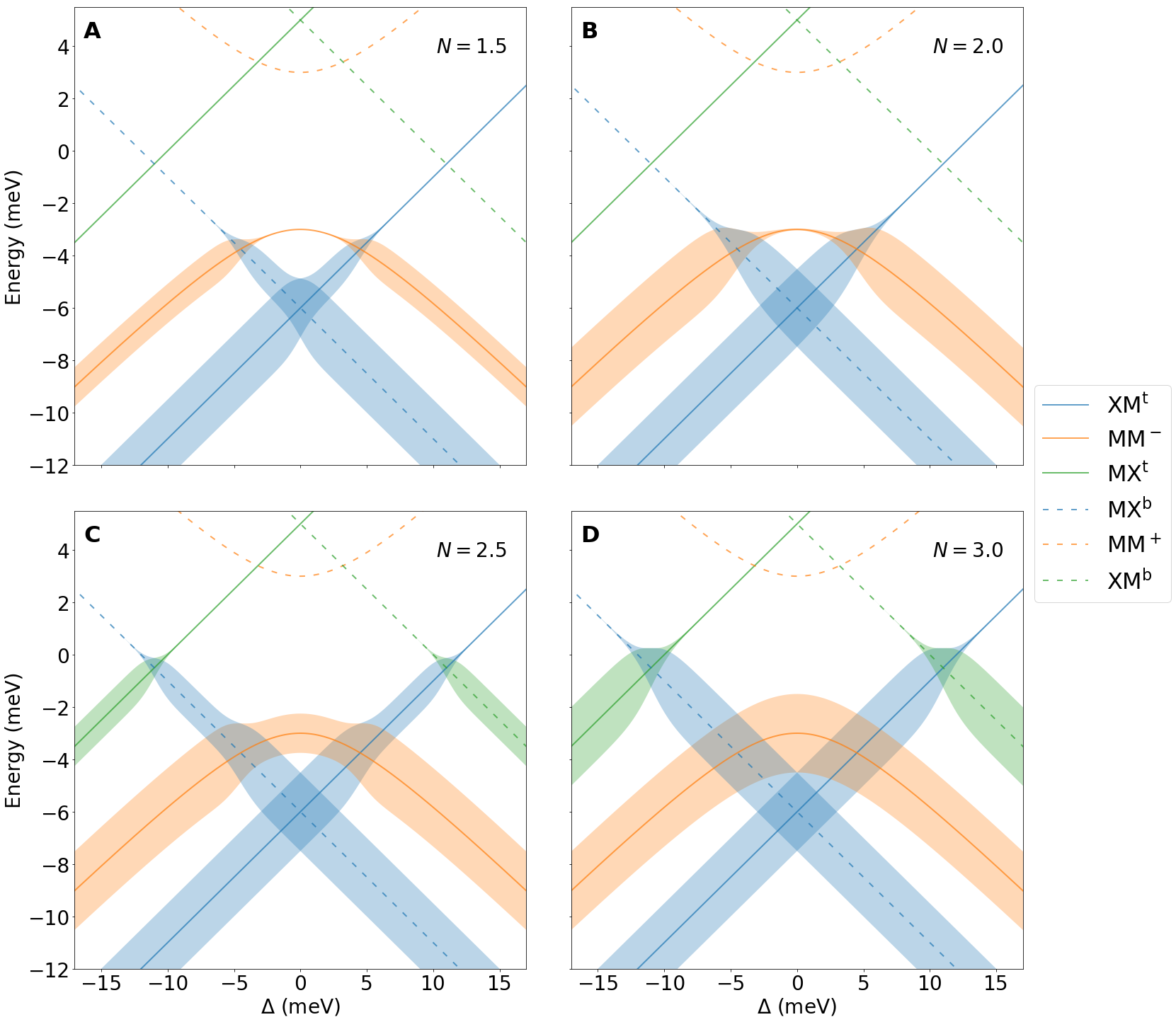}
\caption{{\bf Model calculated ground state energy levels and occupancy.}
        {({\bf A-D})} Calculated hole energy levels as function of energy detuning between top and bottom layers. 
        The width of the semitransparent lines represent hole occupancy at $\nu=1.5$, $\nu=2$, $\nu=2.5$ and $\nu=3$, respectively.
} \label{figS6}
\end{figure*}

We use our experimental observations in order to determine the model parameters.
We find the inter-layer tunnel coupling strength $2t=6.0\pm 0.1\,\rm meV$ from the signatures of the avoided crossing in PL (Figs.~\ref{figS3}A and S3B).
In addition, Figs.~\ref{figS4}A and S4B show \DR\ at $\nu=2$ and $\nu=3$ as function of \VE\ (bottom x axis) and energy detuning (top x axis) between the top and bottom layers. A rough calibration between \VE\ and $\Delta$ can be achieved by using the energy shift that we previously measured for the interlayer exciton~\cite{Shimazaki2020} ($\Delta=V_{\rm E} \cdot 13.8\,\rm meV/V$) while both layers were neutral and assuming that the conversion ratio does not change in the low hole doping regime.
We emphasize that we cannot accurately determine the parameters $\delta_{XM}$ and $\delta_{MX}$ from our experiments; however, a rough estimate can be made from the gap in the avoided crossing at $\nu=2$. We estimate the gap to be $4\pm 2\,\rm meV$ which yields $\delta_{XM}=-6\,\rm meV$. 
Two experimental facts set a lower bound $\delta_{XM}<-3\,\rm meV$: these are (i) coupling strength is $t=3\,\rm meV$ in the $\rm MM$ site, and (ii) the lower energy state for top layer hole is $\rm XM^t$ and that this is the only occupied site at $\nu\leq 1$.
We estimate the crossing point between $\rm XM^t$ and $\rm XM^b$ to be $11\pm 3\,\rm meV$: more specifically, we find the crossing point to be at $10\pm 2\,\rm meV$ from the top layer AP and $12\pm 2\,\rm meV$ from the bottom layer AP. We use this estimate to set $\delta_{MX}=5\,\rm meV$.
The results of the ground state manifold energy eigenstates as function of the energy detuning $\Delta$ between the layers are presented in Fig.~\ref{figS6}.
On the positive detuning side we notice that the lowest energy state is in the $\rm XM^t$\ site, while the second energy state is the $\rm MX^b$\ for $|\Delta|\lesssim 4.2\,\rm meV$ and the $\rm MM^-$\ state for $|\Delta| \gtrsim 4.2\,\rm meV$. The width of semitransparent lines show hole occupancy at a temperature of ${\rm T} = 4.2\,\rm K$.

\begin{figure*}[h!]
\centering
\includegraphics[width=1\textwidth]{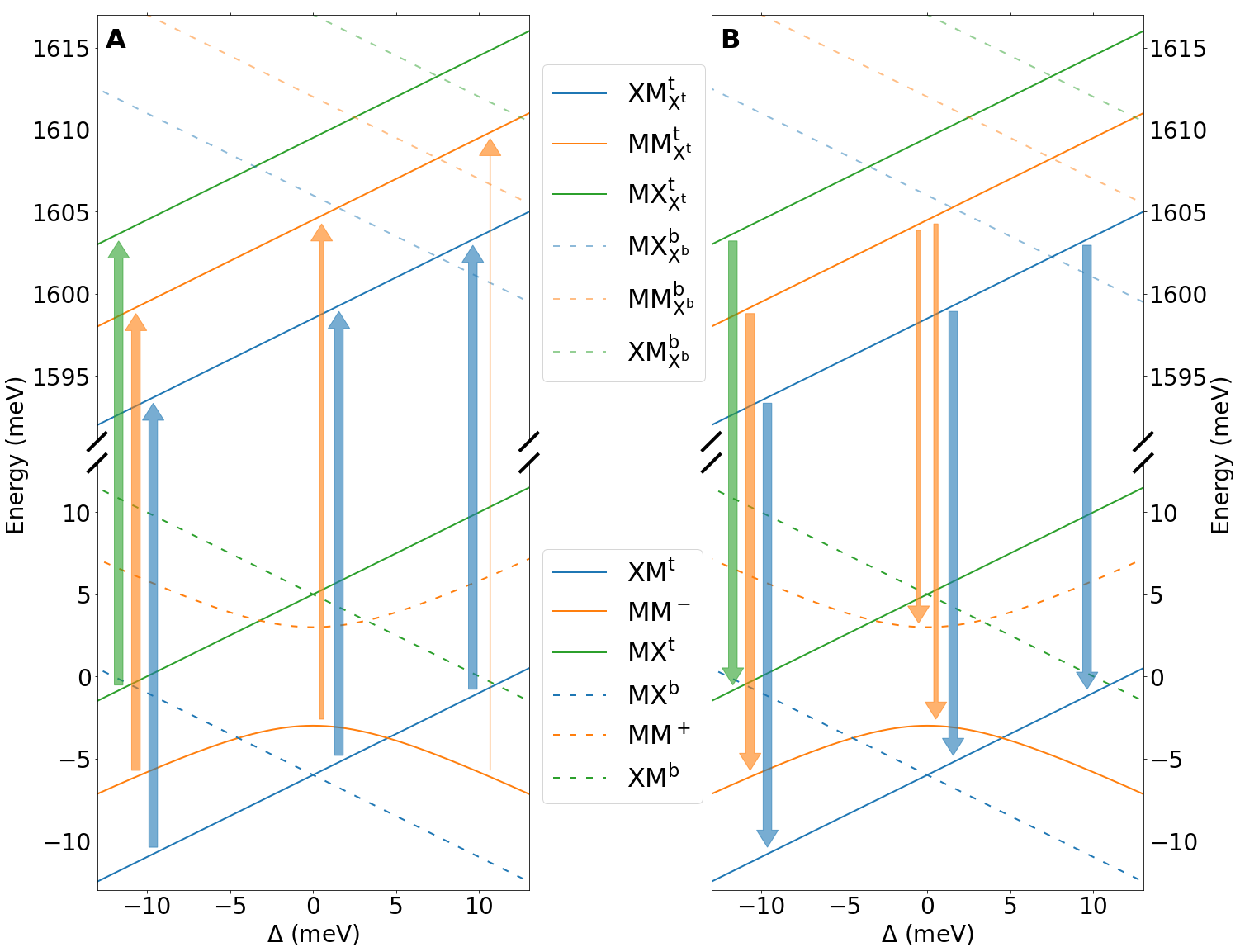}
\caption{{\bf Optical transitions.}
        {({\bf A})} Allowed absorption lines between the ground state manifold to the top layer AP manifold ($\rm AP_{top}$) represented by upward vertical arrows at three $\Delta$ values. Only lines originating from the lowest three energy states at a given $\Delta$ are presented. The width of the arrows represent the oscillator strength of the optical transitions and the arrows color match the initial and final states of the transition. The actual absorption amplitude further depends on the level occupancy (Fig.~\ref{figS6}).
        {({\bf B})} Allowed emission lines between the top layer AP manifold to the ground state manifold represented by downward vertical arrows. Only lines originating from the lowest three AP energy states at a given $\Delta$ are presented.
} \label{figS5}
\end{figure*}

Figure ~\ref{figS5}A show the allowed optical transitions between the ground state manifold to the top layer AP manifold, where only lines that originate from the three lowest energy states are presented. For $\nu=3$ and ${\rm T} = 4.2\,\rm K$, only the marked transitions will be observable. The transitions are represented by upward vertical arrows, while the oscillator strength corresponds to the width of the arrows. At a large negative energy detuning ($\Delta<-11\,\rm meV$) the lowest three states are $\rm XM^t$, $\rm MX^t$ and $\rm MM^-$ where the holes in the $\rm MM^-$ site reside mostly in the top layer. Therefore, the three absorption lines have similar oscillator strength and (nearly) identical transition energies. At a detuning of $\Delta\sim 0$ the three lowest energy states are $\rm XM^t$, $\rm MX^b$ and $\rm MM^-$ from which only two absorption lines to the $\rm AP_{top}$ exists. The holes in the $\rm MM^-$ site have equal probability to occupy the top and bottom layers, resulting in $50\%$ reduction of the oscillator strength for the line that originate from this state to the top layer AP. More importantly, this transition exhibits a blue shift, which originates from the fact that only the ground-state holes exhibit hybridization: even though coherent hole tunneling leads to a lowering of the energy of holes in $\rm MM^-$ sites, the final state of the optical transition is the top layer AP state localized around the $\rm MM^t$ sites. At a larger detuning of $4.2<\Delta<11\,\rm meV$ the lowest three states are the same as for $\Delta\sim 0$; however, the line that originate from the $\rm MM^-$ state continues to blue shift and lose oscillator strength due to its increasing bottom layer weight. Figure ~\ref{figS5}B shows the allowed photoluminescence transitions between the top layer AP manifold to the ground state manifold, where only lines originating from the lowest three $\rm AP$ energy states are presented. At a detuning of $\Delta\sim 0$ the three lowest energy hole states are those localized around the $\rm XM^t$, $\rm MX^b$ and $\rm MM^t$ sites, resulting in three emission lines. The first line corresponds to a transition whose final state is a hole at the $\rm XM^t$ site. The initial state of the other two lines is an AP localized around the $\rm MM^t$ sites, while the two final states are holes residing at the $\rm MM^-$ and $\rm MM^+$ sites.

\clearpage

\section{F\lowercase{eshbach resonance between bottom layer exciton and a top layer hole}} \label{S3}
\begin{figure*}[h!]
\centering
\includegraphics[width=1\textwidth]{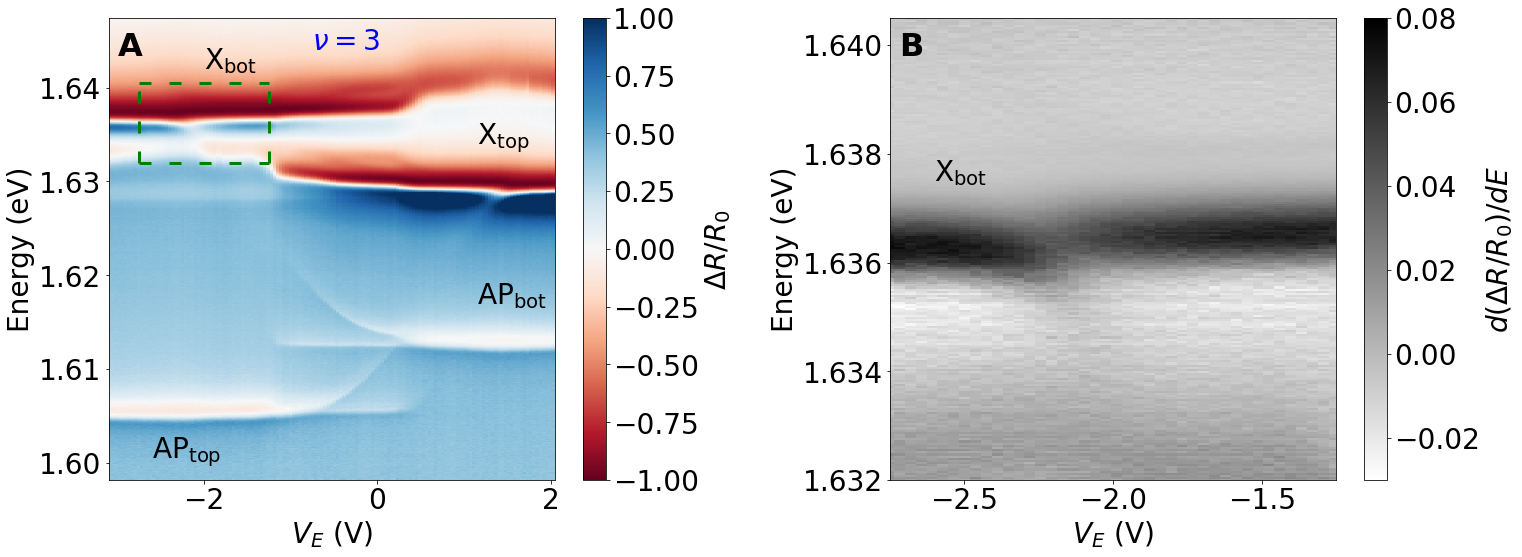}
\caption{{\bf Feshbach resonance.}
        {({\bf A})} $V_{\rm E}$ dependent differential reflectance spectra at a fixed chemical potential for $\nu=3$. Top (bottom) layer exciton/RP is labeled $\text{X}_{\text{top}}$ ($\text{X}_{\text{bot}}$) and top (bottom) layer AP is labeled $\text{AP}_{\text{top}}$ ($\text{AP}_{\text{bot}}$).
        {({\bf B})} $V_{\rm E}$ dependent differential reflectance spectra differentiated with respect to energy E of the green (bottom exciton) dashed box in ({\bf A}).
} \label{figS7}
\end{figure*}

In the main text, we described the Feshbach resonance between the top-layer trion and a scattering state consisting of a top layer exciton and a bottom layer hole. Here we show the complementary Feshbach resonance between the bottom layer trion and a scattering state consisting of bottom layer exciton and a hole in the top layer.
Figure~\ref{figS7}A shows \VE\ dependent \DR\ at $\nu=3$ as in Fig. 4C in the main text and Fig.~\ref{figS7}B shows the derivative of \DR\ with respect to energy in the spectral region highlighted with green (bottom layer exciton) dashed box in Fig.~\ref{figS7}A.

\vspace{1 cm}

\bibliography{References}

\newpage

\clearpage